\documentclass[fleqn,11pt]{article}

\usepackage{amsmath,amsfonts,amssymb}
\usepackage{color}
\usepackage{geometry}
\newtheorem{theorem}{Theorem}
\newtheorem{lemma}[theorem]{Lemma}

\newtheorem{definition}{Definition}

\newcommand{\R}{\mathbb{R}}

\newcommand{\N}{\mathbb{N}}

\newcommand{\cqfd}
{%
\mbox{}%
\nolinebreak%
\hfill%
\rule{2mm}{2mm}%
\newline
\newline
}

\title{On a Boltzmann equation for Haldane statistics.}
\author{Leif ARKERYD $^3$  and Anne NOURI $^{4}$}

\date{}

\begin{document}

\maketitle

{\noindent \bf Abstract.}\hspace{0.1in}
The  study of quantum quasi-particles at low temperatures including their statistics, is a frontier area in modern physics. In a seminal paper Haldane \cite{H}  proposed a definition based on a generalization of the Pauli exclusion principle for fractional quantum statistics.
The present paper is a study of  quantum quasi-particles obeying Haldane statistics in
a fully non-linear kinetic Boltzmann equation model with large initial data on a torus. Strong $L^1$ solutions are obtained for the Cauchy problem. The main results  concern existence, uniqueness and stabililty. Depending on the space dimension and the collision kernel, the results obtained are local or global in time.

\footnotetext[2]{Key words; anyon, Haldane statistics, low temperature kinetic theory, quantum Boltzmann equation.}
\footnotetext[3]{ Mathematical Sciences, 41296 G\"oteborg, Sweden.}
\footnotetext[4]{ Aix-Marseille University, CNRS, Centrale Marseille, I2M UMR 7373, 13453 Marseille, France.}
\section{Haldane statistics and the Boltzmann equation.}
In a previous paper \cite{AN1}, we studied the Cauchy problem for a space-dependent anyon Boltzmann equation \cite{BBM},
\begin{align*}
&\partial _tf(t,x,v)+v_1\partial _xf(t,x,v)= Q_\alpha (f)(t,x,v),\quad t\in \R _+, \hspace*{0.03in} x\in [ 0,1] , \hspace*{0.03in} v=(v_1,v_2) \in \R ^2,\\
&f(0,x ,v )= f_0(x,v).
\end{align*}
The collision operator $Q_\alpha $ in \cite{AN1} depends on a parameter $\alpha \in ] 0,1[ $, and is given by
\begin{eqnarray*}
Q_\alpha (f)(v)= \int _{\R ^2\times S^1}B(\mid v-v_*\mid ,n)\big( f^\prime f^\prime _*F_\alpha (f)F_\alpha (f_*)-ff_*F_\alpha (f^\prime )F_\alpha (f^\prime _*)\big) dv_*dn,
\end{eqnarray*}
with the kernel $B$ of Maxwellian type, $f^\prime $, $f^\prime _*$, $f$, $f_*$ the values of $f$ at $v^\prime $, $v^\prime _*$, $v$ and $v_*$ respectively, where
\begin{eqnarray*}
v^\prime = v-(v-v_*,n)n,\quad v^\prime _*= v_*+(v-v_*,n)n,
\end{eqnarray*}
and the filling factor $F_\alpha $
\begin{eqnarray*}
F_\alpha (f)= (1-\alpha f)^\alpha (1+(1-\alpha )f)^{1-\alpha }.
\end{eqnarray*}
Let us recall the definition of anyon. Consider the wave function $\psi(R,\theta,r,\varphi)$ for two identical particles with center of mass coordinates $(R,\theta)$ and relative coordinates $(r,\varphi)$. Exchanging them,
$\varphi\rightarrow \varphi+\pi$, gives a phase factor $e^{2\pi i}$  for bosons and $e^{\pi i}$ for fermions. In three or more dimensions those are all possibilities. Leinaas and Myrheim proved in 1977 \cite{LM}, that in one and two dimensions any phase factor is
possible in the particle exchange. This became an important topic after the first experimental confirmations in the early 1980-ies, and Frank Wilczek in analogy with the terms bos(e)-ons and fermi-ons coined the name any-ons  for the new quasi-particles with any phase.\\
By moving from spin to a definition in terms of a generalized Pauli exclusion principle, Haldane \cite{H} extended this  to a fractional exclusion statistics valid for any dimension. The conventional Bose-Einstein and Fermi-Dirac statistics are commonly associated with integer spin bosonic elementary particles resp. half integer spin fermionic elementary  particles, whereas the Haldane fractional statistics  is connected with quasi-particles corresponding to elementary excitations in many-body interacting quantum systems.\\
In this paper we consider the Cauchy problem associated to the Boltzmann equation in a torus $[0,1] ^k$, $k\in \{1,2,3\}$, for quantum particles obeying the Haldane statistics;
\begin{align}
&\partial _tf(t,x,v)+\bar{v}\cdot \nabla _xf(t,x,v)= Q(f)(t,x,v),\hspace*{0.03in}(t, x, v)\in \R _+ \times [ 0,1] ^k\times \R ^3 ,\hspace*{0.03in}  v=(v_1,v_2,v_3) \in \R ^3, \label{f} \\
&f(0,x,v)= f_0(x,v),\label{init}
\end{align}
where
\begin{align*}
&\bar{v}= (v_1)\hspace*{0.03in} (\text{  resp.  } \bar{v}= (v_1, v_2), \text{  resp.  } \bar{v}= v) \text{  for  }k=1\hspace*{0.03in} (\text{  resp.  } k=2, \text{  resp.  }k= 3).
\end{align*}
The collision operator $Q$ is given by
\begin{eqnarray*}
Q(f)(v)= \int _{\R ^3\times S^2}B(\mid v-v_*\mid ,n)\big( f^\prime f^\prime _*F_\alpha (f)F_\alpha (f_*)-ff_*F_\alpha (f^\prime )F_\alpha (f^\prime _*)\big) dv_*dn,\quad v\in \R ^3.
\end{eqnarray*}
Strong solutions to the space-homogeneous case were obtained in \cite{A1} for any dimension bigger than one in velocity. Strong solutions to the space-inhomogeneous case were obtained in \cite{AN1} in a periodic slab for two-dimensional velocities. There the proof depends on the two-dimensional velocities setting. In the present paper we prove local in time well-posedness of the Cauchy problem for $k=1$ and collision kernels similar to those used in \cite{AN1}, and for $k\in \{1,2,3\}$ global in time  well-posedness under the supplementary assumption of very soft potential at infinity \cite{V}. The solutions conserve mass, momentum and energy.
\setcounter{equation}{0}
\setcounter{theorem}{0}
\setcounter{equation}{0}
\setcounter{theorem}{0}
%
%
\section{The main results.}
{With $cos\theta = n\cdot\frac{v-v_*}{|v-v_*|}$, the kernel $B(|v-v_*|, n)$ will from now on be written $B(|v-v_*|, \theta )$ and be assumed measurable} with
\begin{equation}\label{hyp1-B}
0\leq B\leq B_0,
\end{equation}
for some $B_0>0$. It is also assumed for some $\gamma, \gamma^\prime >0$, that
\begin{equation}\label{hyp2-B}
B(|v-v_*|, \theta )=0 \hspace*{0.05in}\text{if either}\hspace*{0.05in}   |\cos \theta |<\gamma ^\prime ,\quad
\text{or}\hspace*{0.05in}  1-|\cos \theta|<\gamma ^\prime ,\quad  \text{or  } |v-v_*|< \gamma,
\end{equation}
together with the existence for any $\Gamma>0$ of a constant $c_\Gamma >0$ such that
\begin{equation}\label{hyp3-B}
\int \inf _{u\in [\gamma ,\Gamma ]}B(u, \theta )dn\geq c_\Gamma .
\end{equation}
The initial datum $f_0(x,v)$, 
\begin{align}
& \text{periodic in }\hspace*{0.02in}x, \hspace*{0.03in}\text{is a measurable function with values in}\hspace*{0.04in} [0,\frac{1}{\alpha }] ,\label{Hyp5-f0}
\end{align}
and such that for some positive constants $c_0$ and $\tilde{c}_0$,
\begin{align}
&(1+|v|^2)f_0(x,v) \in L^1([ 0,1] ^k\times \R ^3), \label{Hyp1-f0}\\
&\int \sup_{x\in[0,1]^k} f_0(x,v)dv= c_0 ,\label{Hyp2-f0}\\
&\int \sup_{x\in[0,1]^k} |v|^2f_0(x,v)dv= \tilde{c}_0 ,\label{Hyp3-f0}\\
&\text{for any subset $X$ of $\R ^3 $ of positive measure},\quad \int _X\hspace*{0.04in}\inf _{x\in [0,1] ^k}f_0(x,v)dv>0. \label{Hyp4-f0}
\end{align}
Denote by
\begin{equation}\label{f-sharp}
f^{\sharp }(t,x,v)= f(t, x+t\bar{v}, v)\quad (t,x,v)\in \R _+\times [ 0,1] ^k\times \R ^3,\quad \bar{v}= (v_1,\cdot \cdot \cdot , v_k)\in \R ^k.
\end{equation}
Strong solutions to the Cauchy problem with initial value $f_0$ associated to the quantum Boltzmann equation (\ref{f}) are considered in the following sense.
\begin{definition}\label{strong-solution}
$f$ is a strong solution to (\ref{f}) on the time interval $I$ if
\begin{eqnarray*}
f\in\mathcal{C}^1(I;L^1([0,1] ^k\times \R^3)),
\end{eqnarray*}
and
\begin{equation}\label{eq-along-characteristics}
\frac{d}{dt}f^{\sharp }= \big( Q(f)\big) ^{\sharp },\quad \text{on   } I\times [ 0,1] ^k\times \R ^3.
\end{equation}
\end{definition}
The main results of the present paper are given in the following theorems.\\
\setcounter{theorem}{0}
%
%
\begin{theorem}\label{th-local}
\hspace*{0.1in}\\
Under the assumptions (\ref{hyp1-B})-(\ref{Hyp2-f0}) and (\ref{Hyp4-f0}), there is a time $T_0>0$, so that there exists a unique periodic in $x$, strong solution $f\in\mathcal{C}^1([0,T_0[ ;L^1([0,1]\times\R^3))$ of (\ref{f})-(\ref{init}). It depends continuously in $\mathcal{C}([0,T_0[;L^1([0,1]\times\R^3))$ on the initial $L^1$-datum. It conserves mass, momentum and energy.
\end{theorem}
%
%
\begin{theorem}\label{th-global}
\hspace*{0.1in}\\
Under the assumptions (\ref{hyp1-B})-(\ref{Hyp4-f0}) and the supplementary assumption of very soft collision kernels at infinity,
 \begin{equation}\label{hyp4-B}
B(u, \theta )= B_1(u)B_2(\theta )\quad \text{with   }|B_1(u)|\leq c|u|^{-3-\eta }\text{   for some  }\eta >0, \text{and  }B_2\text{   bounded},
\end{equation}
there exists a unique periodic in $x$, strong solution $f\in\mathcal{C}^1([0,\infty [;L^1([0,1] ^k\times\R^3))$ of (\ref{f})-(\ref{init}) \\
for $k\in \{1, 2, 3\} $. For any $T>0$ it continuously depends in $\mathcal{C}([0,T];L^1([0,1] ^k\times \R^3))$ on the initial $L^1$-datum. It conserves mass, momentum and energy.
\end{theorem}
%
%
\underline{\bf Remarks.}\\
Theorem \ref{th-local} is restricted to the slab case, since its proof below uses an estimate for the Bony functional only valid in one space dimension. \\
Theorems \ref{th-local} and \ref{th-global} also hold with the same proofs in the fermion case where $\alpha=1$, in particular giving strong solutions to the Fermi-Dirac equation.\\
Theorems \ref{th-local} and \ref{th-global} also hold with a limit procedure when $\alpha \rightarrow 0$ in the boson case where $\alpha= 0$, in particular giving strong solutions to the Boltzmann Nordheim equation \cite{N}. It is the object of a separate paper \cite{AN3} (see also \cite{EV}, \cite{Lu4} and \cite{BE})\\
Theorems \ref{th-local} and \ref{th-global} also hold for $v\in \R ^n$, $n\geq 3$.\\
The proofs in \cite{AN1} strongly rely on the property that for any unit vector $n$ with direct orthogonal unit vector $n_\perp $, either $n_1$ or $n_{\perp 1}$  is bigger that $\frac{1}{\sqrt{2}}$, where $n_1$ (resp. $n_{\perp 1}$) is the component of $n$ (resp. $n_\perp $) along the $x$- axis. This allows to control the mass density of the solution from its Bony functional. This is no more the case in the three-dimensional velocity setting of the present paper. 
It is why our results are local in time under the same assumptions on the collision kernel $B$ as in \cite{AN1}. They are global in time under the supplementary assumption of a very soft potential at infinity.\\
\hspace*{1.in}\\
The paper is organized as follows. Approximations are introduced in Section \ref{approx-bony} for $k\in \{ 1,2,3\} $ together with for $k=1$, a control of their Bony functional. Their mass density is uniformly controlled under the assumptions of Theorem \ref{th-local} (resp. Theorem \ref{th-global}) in Section \ref{mass-density-local} (resp. Section \ref{mass-density-global}). The well-posedness of the Cauchy problem is proven in Section \ref{pf-th}. Conservation of mass, momentum and energy is proven in Section \ref{conservations}.

\setcounter{equation}{0}
\setcounter{theorem}{0}
%
%
\section{Preliminaries on solution approximations and the Bony functional.}
\label{approx-bony}
\setcounter{equation}{0}
\setcounter{theorem}{0}
Let $k\in \{1, 2, 3\} $. For any $j\in \N ^*$, denote by $\psi_j$, the cut-off function with
\[ \begin{aligned}
&\psi_j(r)=0\quad \text{if   }r>j^2
&\text{and}\quad \psi_j(r)= 1\quad \text{if  }
r\leq j^2,
\end{aligned}\]
and set
\[ \begin{aligned}\chi_j(v,v_*)=\psi_j(|v ^2| +|v_*| ^2 ) .
\end{aligned}\]
Let $F_j$ be the $C^1$ function defined on $[ 0,\frac{1}{\alpha }] $ by
\begin{eqnarray*}
 F_j(y)= \frac{1-\alpha y}{(\frac{1}{j} +1-\alpha y)^{1-\alpha }} (1+(1-\alpha )y)^{1-\alpha }.
\end{eqnarray*}
Denote by $Q_j$ (resp. $Q_j^+$, {and $Q_j^-$ to be used later}), the operator
\begin{align}
&Q_j(f)(v):= \int B(|v-v_*|,{\theta} )\chi _j(v,v_*)\Big( f^\prime f^\prime _*F_j(f)F_j(f_*)-ff_*F_j(f^\prime )F_j(f^\prime _*)\Big) dv_*dn  ,\nonumber \\
&(\text{resp. its gain part    }Q_j^+(f)(v):= \int B(|v-v_*|,{\theta } )\chi _j(v,v_*)f^\prime f^\prime _*F_j(f)F_j(f_*)dv_*dn ,\label{Qj+} \\
&{\text{and its loss part    }Q_j^-(f)(v):= \int B(|v-v_*|,{\theta } )\chi _j(v,v_*)ff_*F_j(f^\prime )F_j(f^\prime _*)dv_*dn \hspace*{0.02in} ).}\label{Qj-}
\end{align}
For $j \in \N ^*$, let a mollifier $\varphi _j$ be defined by  $\varphi _j(x,v)= j^{3+k}\varphi (jx,jv)$, where
\begin{align*}
&\varphi \in C_0^\infty (\R ^{3+k}),\quad support (\varphi )\subset [ 0,1] ^k\times \{ v\in \R ^3;\lvert v\rvert \leq 1\},\\
& \varphi \geq 0,\quad\int _{[0,1] ^k\times \R ^3}\varphi (x,v)dxdv= 1.
\end{align*}
Let
\begin{equation}\label{df-init-approx}
f_{0,j} \text{  be the restriction to  }[ 0,1] ^k\times \{v; \lvert v\rvert \leq j\} \text{  of   }\big( \min \{f_0, \frac{1}{\alpha }-\frac{1}{j} \} \big) \ast \varphi _j.
\end{equation}
The following lemma concerns a corresponding approximation of (\ref{f})-(\ref{init}) for  $k\in \{ 1, 2, 3\} $.\\
%
%
\begin{lemma}
\hspace{1cm}\\ 
For any $T>0$, there is a unique solution $f_j\in {C}\big( [ 0,T] \times [ 0,1] ^k;L^1( \{v; \lvert v\rvert \leq j\} )\big) $ to
\begin{equation}\label{eq-f_j}
\partial _tf_j+\bar{v}\cdot \nabla _xf_j= Q_j(f_j),\quad f_j(0,\cdot ,\cdot )= f_{0,j}.
\end{equation}
There is $\eta _j>0$ such that $f_j$ takes its values in $ [ 0,\frac{1}{\alpha }-\eta _j] $.\\
The solution conserves mass, momentum and energy.
\end{lemma}
\underline{Proof of Lemma 3.1.}\\
Let $T>0$ be given. We shall first prove by contraction that for $T_1>0$ and small enough, there is a unique solution
\begin{eqnarray*}
f_{j}\in C\big( [ 0,T_1] \times [ 0,1] ^k; L^1( \{ v; \lvert v\rvert \leq j\} )\big) \cap \{f; f\in [ 0,\frac{1}{\alpha } ] \}
\end{eqnarray*}
to (\ref{eq-f_j}). Let the map $\mathcal{C}$ be defined on {periodic in $x$ functions in}
\begin{eqnarray*}
C \big( [ 0,T] \times [ 0,1] ^k;L^1( \{ v; \lvert v\rvert \leq j\} )\big) \cap \{f; f\in [ 0,\frac{1}{\alpha } ] \}
\end{eqnarray*}
by $\mathcal{C}(f)= g$, where
\[ \begin{aligned}
&\partial _tg +\bar{v}\cdot \nabla _xg = (1-\alpha g)\Big( \frac{1+(1-\alpha )f}{\frac{1}{j}+1-\alpha f}\Big) ^{1-\alpha }\int B\chi _j{f}^\prime {f}^\prime _*F_j({f}_*)dv_*dn \\
&\hspace*{1.in}-g\int B\chi _j{f}_*F_j(f^\prime )F_j(f^\prime _*)dv_*dn  ,\\
&g(0,\cdot ,\cdot )= f_{0,j}.
\end{aligned}\]
The previous linear partial differential equation has a unique periodic solution
\begin{eqnarray*}
g\in C ([ 0,T] \times [ 0,1] ^k;L^1( \{ v; \lvert v\rvert \leq j\} )) .
\end{eqnarray*}
For $f$ with values in $ [ 0,\frac{1}{\alpha } ] $, $g$ takes its values in $ [ 0,\frac{1}{\alpha } ] $. Indeed, denoting by
\[ \begin{aligned}
&\bar{\sigma }_f:= \alpha \Big( \frac{1+(1-\alpha )f)}{\frac{1}{j}+1-\alpha f}\Big) ^{1-\alpha }\int B\chi _jf^\prime f^\prime _*F_j(f_*)dv_*dn + \int B\chi _jf_*{F}_{j}(f^\prime ){F}_{j}(f^\prime _*)dv_*dn  ,
\end{aligned}\]
and
\begin{eqnarray*}
g^\sharp (t,x,v)= g(t,x+t\bar{v},v),
\end{eqnarray*}
it holds that 
\[ \begin{aligned}
g^\sharp (t,x,v)&= f_{0,j}(x,v)e^{-\int _0^t\bar{\sigma }_f^\sharp (r,x,v)dr}\\
&+\int_0^tds\Big( \big( \frac{1+(1-\alpha )f}{\frac{1}{j}+1-\alpha f}\big) ^{1-\alpha }\int B\chi _j{f}^\prime {f}^\prime _*F_j({f}_*)dv_*dn\Big)^\sharp (s,x,v)e^{-\int _s^t\bar{\sigma }_f^\sharp (r,x,v)dr}\\
&\geq f_{0,j}(x,v)e^{-\int _0^t\bar{\sigma }_f^\sharp (r,x,v)dr}\geq 0,
\end{aligned}\]
and
\[ \begin{aligned}
(1-\alpha g)^\sharp (t,x,v)&= (1-\alpha f_{0,j})(x,v)e^{-\int _0^t\bar{\sigma }_f^\sharp (r,x,v)dr}\\
&+\int _0^t\Big( \int B\chi _jf_*F_j(f^\prime)F_j(f^\prime _*)dv_*dn\Big) ^\sharp (s,x,v)e^{-\int _s^t\bar{\sigma }_f^\sharp (r,x,v)dr}ds\\
&\geq (1-\alpha f_{0,j})(x,v)e^{-\int _0^t\bar{\sigma }_f^\sharp (r,x,v)dr}\geq 0.
\end{aligned}\]
$\mathcal{C}$ is a contraction on $C([0,T_1] \times [ 0,1] ^k; L^1(\{ v;\lvert v\rvert \leq j\} ))\cap \{f; f\in [ 0,\frac{1}{\alpha } ] \}$, for $T_1>0$ small enough only depending on $j$, since the derivative of the map $F_j$ is bounded by $(3j\alpha ^{\alpha -1}+1)j^{1-\alpha}$ on $[ 0,\frac{1}{\alpha }] $. Let $f_j$ be its fixed point, i.e. the solution of (\ref{eq-f_j}) on $[ 0,T_1] $. The argument can be repeated and the solution continued up to $t=T$. By Duhamel's form for $f_j$ (resp. $1-\alpha f_j$),
\[ \begin{aligned}
f_j^\sharp (t,x,v)&\geq f_{0,j}(x,v)e^{-\int _0^t\bar{\sigma }_{f_j}^\sharp (r,x,v)dr}\geq 0,
\quad (t, x)\in [ 0,T] \times [ 0,1] ^k,\hspace{0.03in}\lvert v\rvert \leq j,
\end{aligned}\]
(resp.
\[ \begin{aligned}
(1-\alpha f_j)^\sharp (t,x,v)&\geq (1-\alpha f_{0,j})(x,v)e^{-\int _0^t\bar{\sigma }_{f_j}^\sharp (r,x,v)dr}\\
&\geq \frac{1}{je^{cj^4T}},
\quad (t, x)\in [ 0,T]\times [ 0,1] ^k,\hspace{0.03in}\lvert v\rvert \leq j).
\end{aligned}\]
Consequently, for some $\eta_j>0$, there  is a {periodic in $x$} solution
\begin{eqnarray*}
f_j\in C([ 0,T] \times [ 0,1] ^k;L^1( \{ v;\lvert v\rvert \leq j\} ))
\end{eqnarray*}
to (\ref{eq-f_j}) with values in $[ 0, \frac{1}{\alpha }-{\eta_j}]$.\\
If there were another nonnegative local solution $\tilde{f}_{j}$ to (\ref{eq-f_j}), defined on $[ 0,T^\prime ] $ for some $T^\prime \in ] 0,T] $, then by the exponential form it would strictly stay below $\frac{1}{\alpha}$. The difference $f_j-\tilde{f}_j$ would for some constant $c_{T^\prime}$ satisfy
\begin{eqnarray*}
\int \lvert (f_j -\tilde{f}_j)^\sharp (t,x,v)\rvert dxdv\leq c_{T^\prime }\int _0^t\lvert (f_j-\tilde{f}_j)^\sharp (s,x,v)\rvert dsdxdv,\hspace*{0.03in}t\in [ 0,T^\prime ],\quad  (f_j-\tilde{f}_j)^\sharp (0,x,v)= 0,
\end{eqnarray*}
implying that the difference {would be} identically zero on $[ 0,T^\prime ] $. Thus $f_j$ is the unique solution on $[ 0,T] $ to (\ref{eq-f_j}), and has its range contained in $[ 0,\frac{1}{\alpha }-\eta _j] $. 
 \cqfd
Denote by $M_j$ the mass density
\begin{equation}\label{df-mass-density}
M_j(t)= \int \sup _{(s,x)\in [ 0,t] \times [ 0,1]}f^\sharp _j(s,x,v)dv.
\end{equation}
In Lemma \ref{mass-tails} the tails for large velocities of the mass are controlled with respect to the mass density. 
%
%
\begin{lemma}\label{mass-tails}
\hspace*{0.02in}\\
Given $T>0$, the solution $f_j$ of (\ref{eq-f_j}) satisfies
\begin{eqnarray*}
\int _0^1\int_{|v|>\lambda} |v|\sup_{t\in [ 0,T] }f_j^\sharp (t,x,v)dvdx\leq\frac{c_T}{\lambda}M_j(T),\quad j\in \N ,
\end{eqnarray*}
where $c_T$ only depends on $T$ and $\int |v|^2f_0(x,v)dxdv$.
\end{lemma}
\underline{Proof of Lemma \ref{mass-tails}.} \\
Denote $f_j$ by $f$ for simplicity. By the non-negativity of $f$,
\begin{equation}
 \sup_{t\in [ 0,T] }f^\sharp (t,x,v)\leq  f_0(x,v)+\int_0^T(Q_j^+(f))^\sharp (s,x,v)ds,\label{control-by-gain}
\end{equation}
where $Q_j^+(f)$ is defined in (\ref{Qj+}). Integration with respect to $(x,v)$  for $|v|>\lambda$, gives
\begin{align*}
&\int _0^1\int_{|v|>\lambda}|v|\sup_{t\in [ 0,T] }f^\sharp (t,x,v)dvdx
\leq  \int \int_{|v|>\lambda}|v|f_0(x,v)dvdx+\int_0^T\int_{|v|>\lambda} B{\chi}_j\\
&|v| f(s,x+sv_1,v^\prime )f(s,x+sv_1,v^\prime _*)F_j(f)(s,x+sv_1,v)F_j(f)(s,x+sv_1,v_*)dvdv_*dndxds.
\end{align*}
Here in the last integral, either $|v^\prime |$ or $|v^\prime _*|$ is the largest and larger than $\frac{\lambda}{\sqrt 2}$. The two cases are symmetric, and we discuss the case $|v^\prime |\geq|v^\prime _*|$. After a translation in $x$, the integrand of the r.h.s of the former inequality is estimated {from above} by
\begin{eqnarray*}
c |v^\prime |f^{\#} (s,x,v^\prime )\sup_{(t,x)\in [ 0,T] \times [0,1]}f^{\#}(t,x,v^\prime _*).
\end{eqnarray*}
The change of variables {$(v,v_*,n)\rightarrow (v^\prime ,v^\prime _*,-n)$} and the integration over
 \begin{eqnarray*}
(s,x,v,v_*,n)\in [ 0,T] \times [ 0,1] \times \{ v\in \R ^3; |v| >\frac{\lambda}{\sqrt 2}\} \times \R ^3 \times \mathcal{S}^2,
\end{eqnarray*}
give the bound
\begin{align*}
&\frac{c}{\lambda}\Big( \int_0^T\int |v|^2f^{\#}(s,x,v)dxdvds\Big) \Big( \int \sup_{(t,x)\in [ 0,T] \times [0,1]} f^{\#}(t,x,v_*)dv_*\Big) \\&\leq \frac{cTM_j(T)}{\lambda}\int |v|^2f_0(x,v)dxdv.
\end{align*}
The lemma follows.                        \cqfd
%
%
%
For $k=1$ there is a Bony type inequality available (cf \cite{B} \cite{CI}) as follows.
\begin{lemma}\label{bony}
\hspace*{0.1in}\\
For any $n\in \mathcal{S}^2$, denote by $n_1$ the component of $n$ along the $x$-axis. It holds that
\begin{align}
&\int_0^t\int n_1^2[(v-v_*)\cdot n]^2B{\chi}_jf_jf_{j*}F_j(f^\prime _j)F_j(f^\prime _{j*})dvdv_*dn dxds\leq c'_0(1+t),\quad t>0,\hspace*{0.03in} j\in \N ^*,\label{ineq-bony}
\end{align}
with $c'_0$ only depending on $\int f_0(x,v)dxdv$ and $\int |v|^2f_0(x,v)dxdv$.
\end{lemma}
\underline{Proof of Lemma \ref{bony}.} \\
Denote $f_j$ by $f$. The integral over time of the first component of momentum $\int v_1f(t,0,v)dv$ (resp. $\int v_1^2f(t,0,v)dv$ ) is first controlled. Let $\beta \in C^1([ 0,1] )$ be such that $\beta (0)= -1$ and $\beta (1)= 1$. Multiply (\ref{eq-f_j}) for $k= 1$ by $\beta (x)$ (resp. $v_1\beta (x)$ ) and integrate over $[ 0,t] \times [ 0,1] \times \R ^3$. It gives
\[ \begin{aligned}
\int _0^t\int v_1f(\tau ,0,v)dvd\tau = \frac{1}{2}\big( \int \beta (x)f_{0,j}(x,v)dxdv&-\int \beta (x)f(t,x,v)dxdv\\
&+\int _0^t\int \beta ^\prime (x)v_1f(\tau ,x,v)dxdvd\tau\big) ,
\end{aligned}\]
\Big( resp.
\[ \begin{aligned}
\int _0^t\int v_1^2f(\tau ,0,v)dvd\tau = \frac{1}{2}\big( \int \beta (x)v_1f_{0,j}(x,v)dxdv&-\int \beta (x)v_1f(t,x,v)dxdv\\
&+\int _0^t\int \beta ^\prime (x)v_1^2f(\tau ,x,v)dxdvd\tau\big) \Big) .
\end{aligned}\]
Consequently, using the conservation of mass and energy of $f$,
\begin{align}\label{bony-1}
\lvert \int _0^t\int v_1f(\tau ,0,v)dvd\tau \rvert +\int _0^t\int v_1^2f(\tau ,0,v)dvd\tau \leq c(1+t).
\end{align}
Let
\begin{eqnarray*}
\mathcal{I}(t)= \int _{x<y}(v_1-v_{*1})f(t,x,v)f(t,y,v_*)dxdydvdv_*.
\end{eqnarray*}
It results from
\begin{align*}
\mathcal{I}'(t)=& -\int (v_1-v_{*1})^2f(t,x,v)f(t,x,v_*)dxdvdv_*\\
&+2\int v_{*1}(v_{*1}-v_1)f(t,0,v_*)f(t,x,v)dxdvdv_*,
\end{align*}
and the conservations of the mass, momentum and energy of $f$ that
\begin{align}\label{bony-2}
&\int _0^t \int_0^1 \int (v_1-v_{*1})^2 f(s,x,v)f_*(s,x,v_*)dvdv_*dxds\nonumber \\
&\leq 2\int f_0(x,v)dxdv\int \lvert v_1\rvert f_0(x,v)dv+ 2\int f(t,x,v)dxdv\int \lvert v_1\rvert f(t,x,v)dxdv\nonumber \\
&+2\int _0^t\int v_{*1}(v_{*1}-v_1)f(\tau ,0,v_*)f(\tau ,x,v)dxdvdv_*d\tau \nonumber \\
&\leq 2\int f_0(x,v)dxdv\int (1+\lvert v\rvert ^2)f_0(x,v)dv+ 2\int f(t,x,v)dxdv\int (1+\lvert v\rvert ^2) f(t,x,v)dxdv\nonumber \\
&+2\int _0^t(\int v_{*1}^2f(\tau ,0,v_*)dv_*)d\tau\int f_0(x,v)dxdv\nonumber \\
&-2\int _0^t(\int v_{*1}f(\tau ,0,v_*)dv_*)d\tau\int  v_1f_0(x,v)dxdv\nonumber \\
&\leq c\Big( 1+\int _0^t\int v_1^2f(\tau ,0,v)dvd\tau +\lvert \int _0^t\int v_1f(\tau ,0,v)dv\rvert \Big) . \nonumber
\end{align}
And so, by (\ref{bony-1}),
\begin{equation}\label{bony-3}
\int _0^t \int_0^1 \int (v_1-v_{*1})^2 f(\tau ,x,v)f(\tau ,x,v_*)dxdvdv_*d\tau \leq c(1+t).
\end{equation}
Here, $c$ is a constant depending only on $\int f_0(x,v)dxdv$ and $\int \lvert v\rvert ^2f_0(x,v)dxdv$. \\
Denote by $u_1=\frac{\int v_1fdv}{\int fdv}$. It holds
\begin{align}\label{bony-4}
\int_0^t\int_0^1 \int (v_1-u_1)^2 B{\chi}_jff_*&F_j(f')F_j(f'_*)(s,x,v,v_*,n)dvdv_*dn dxds\nonumber \\
&\leq c\int_0^t  \int_0^1 \int (v_1-u_1)^2 ff_*(s,x,v,v_*)dvdv_* dxds\nonumber \\
&= \frac{c}{2}\int _0^t \int_0^1 \int (v_1-v_{*1})^2 ff_*(s,x,v,v_*)dvdv_*dxds\nonumber \\
&\leq c(1+t).
\end{align}
Multiply equation (\ref{eq-f_j}) for $f$  by $v_1^2$, integrate and use that $\int v_1^2Q_j(f)dv= \int (v_1-u_1)^2Q_j(f)dv$ \\
and (\ref{bony-4}). It results
\[ \begin{aligned}
&\int _0^t\int (v_1-u_1)^2B{\chi}_jf^\prime f^\prime _*F_j(f)F_j(f_*)dvdv_*dn dxds= \int v_1^2f(t,x,v)dxdv\\
&-\int v_1^2f_{0,j}(x,v)dxdv+\int _0^t\int (v_1-u_1)^2B{\chi}_jff_*F_j(f^\prime )F_j(f^\prime _*)dxdvdv_*dn ds\\
&<c'(1+t),
\end{aligned}\]
where $c'$ is a constant only depending on $\int f_0(x,v)dxdv$ and $\int \lvert v\rvert ^2f_0(x,v)dxdv$.\\
\hspace{1cm}\\
After  a change of variables the left hand side can be written
\[ \begin{aligned}
&\int _0^t\int (v'_1-u_1)^2B{\chi}_jff_*F_j(f')F_j(f'_*)dvdv_*dn dxds\\
&= \int _0^t\int (c_1-n_1[(v-v_*)\cdot n])^2B{\chi}_jff_*F_j(f')F_j(f'_*)dvdv_*dn dxds,
\end{aligned}\]
where $c_1=v_1-u_1$.
Expand $(c_1-n_1[(v-v_*)\cdot n])^2$, remove the  positive term containing $c_1^2$. \\
\\
The term containing $n_1^2[(v-v_*)\cdot n]^2$ is estimated as follows;
\hspace*{0.1in}\\
\begin{align*}
&\int_0^t\int n_1^2[(v-v_*)\cdot n]^2B{\chi}_jff_*F_j(f')F_j(f'_*)dvdv_*dn dxds\\
&\leq c^\prime {(1+t)}+2\int _0^t\int (v_1-u_1)n_1[(v-v_*)\cdot n]B{\chi}_jff_*F_j(f')F_j(f'_*)dvdv_*dn dxds\\
&\leq {c^\prime }{(1+t)}+2\int _0^t\int \Big( v_1\sum^3_{l=2}(v_l-v_{*l})n_1n_l
\Big) B{\chi}_jff_*F_j(f')F_j(f'_*)dvdv_*dn dxds,
\end{align*}
since
\[ \begin{aligned}
\int u_1(v_l-v_{*l})n_1n_l{\chi}_jBff_*F_j(f')F_j(f'_*)dvdv_*dn dx
= \hspace*{0.01in}0,\quad\quad l=2, 3,
\end{aligned}\]
by an exchange of the variables $v$ and $v_*$. Moreover, exchanging first the variables $v$ and $v_*$,
\[ \begin{aligned}
2\int _0^t&\int v_1\sum^3_{l=2}(v_l-v_{*l})n_1n_lB{\chi}_jff_*F_j(f')F_j(f'_*)dvdv_*dn dxds\\
= &\int _0^t\int (v_1-v_{*1})\sum^3_{l=2}(v_l-v_{*l})n_1n_lB{\chi}_jff_*F_j(f')F_j(f'_*)dvdv_*dn dxds\\
\leq &\frac{1}{\beta ^2}\int _0^t\int (v_1-v_{*1})^2B{\chi}_jff_*F_j(f')F_j(f'_*)dvdv_*dn dxds\\
&+\frac{{\beta ^2}}{4}\int _0^t\int \sum^3_{l=2}(v_l-v_{*l})^ 2n^2_1n_l^2B{\chi}_jff_*F_j(f')F_j(f'_*)dvdv_*dn dxds\\
\leq &\frac{2c'}{{\beta ^2}}{(1+t)}+\frac{{\beta ^2}}{4}\int _0^t\int n_1^2 \sum^3_{l=2}(v_l-v_{*l})^2n_l^2B{\chi}_jff_*F_j(f')F_j(f'_*)dvdv_*dn dxds,
\end{aligned}\]
for any $\beta >0$. It follows that
\begin{eqnarray*}
\int_0^t\int n_1^2[(v-v_*)\cdot n]^2B{\chi}_jff_*F_j(f')F_j(f'_*)dvdv_*dn dxds\leq c'_0(1+t),
\end{eqnarray*}
with $c'_0$ only depending on $\int f_0(x,v)dxdv$ and $\int \lvert v\rvert ^2f_0(x,v)dxdv$. This completes the proof of the lemma. \cqfd
%
%
%
\section{Control of the mass density under the assumptions of Theorem \ref{th-local}.}\label{mass-density-local}
\setcounter{theorem}{0}
\setcounter{equation}{0}
Let $k= 1$. Lemmas \ref{integral-dxdv-local} to \ref{df-T-local} are devoted to the local in time uniform control with respect to $j$ of the mass density defined in (\ref{df-mass-density}).
%
%
\begin{lemma}\label{integral-dxdv-local}
\hspace*{0.2in}\\
For any $\epsilon>0$, there exists a constant $c'_1$ only depending on
$\int f_0(x,v)dxdv$ and $\int |v|^2f_0(x,v)dxdv$, such that
\begin{equation}\label{dxdv-local}
\int \sup_{s\in [ 0,t] }f_j^\sharp (s,x,v)dxdv\leq c'_1\Big( (1+\frac{1}{\epsilon ^2})(1+t)+\epsilon tM_j(t)\Big) ,\quad t>0,\quad j\in \N ^*.
\end{equation}
\end{lemma}
\hspace*{1.in}\\
\underline{Proof of Lemma \ref{integral-dxdv-local}.} \\
Denote $f_j$ by $f$ for simplicity. By (\ref{control-by-gain}),
\begin{align}
&\sup_{s\in [0,t] }f^\sharp (s,x,v)\leq  f_0(x,v)\nonumber \\
&+\int_0^t\int B{\chi}_j
f(r,x+rv_1,v^\prime )f(r,x+rv_1,v^\prime _*)F_j(f)^\sharp(r,x,v)F_j(f)(r,x+rv_1,v_*)dndv_*dr.\label{control-by-gain-a}
\end{align}
{For any $(v,v_*)\in \R ^3\times \R ^3$, let $\mathcal{N}_\epsilon$ be the set of $n\in \mathcal{S} ^2$ with $\max \{ n_1,n_{\perp 1}\} <\epsilon$, where $n_\perp $ is the unit vector in the direction $v-v^\prime _*$ (orthogonal to $n$) in the plane defined by $v-v_*$ and $n$, and $n_1$ is the component of $n$ along the $x$-axis.}\\
Let $\mathcal{N}^c _\epsilon$ be the complement of $\mathcal{N}_\epsilon$ in $\mathcal{S}^2$.  Denote by
\begin{eqnarray*}
\mathcal{I}_\epsilon (t)= \int_0^t\int \int _{\mathcal{N}_\epsilon} B{\chi}_j
f(r,x+rv_1,v^\prime )f(r,x+rv_1,v^\prime _*)\\
F_j(f)^\sharp(r,x,v)F_j(f)(r,x+rv_1,v_*)dndvdv_*dxdr.
\end{eqnarray*}
(\ref{ineq-bony}) also holds with $n_1$ replaced by $n_{\perp 1}$. Integrating (\ref{control-by-gain-a}) with respect to $(x,v)$ and using (\ref{hyp2-B}) and Lemma \ref{bony} leads to
\begin{align}\label{lemma4.1-1}
&\int \sup_{s\in [0, t]}f^\sharp (s,x,v)dxdv
\leq  \int f_0(x,v)dxdv+\mathcal{I}_\epsilon (t)\nonumber \\
&+\int_0^t\int \int _{\mathcal{N}^c _\epsilon} B{\chi}_j
f(r,x+rv_1,v^\prime )f(r,x+rv_1,v^\prime _*)\nonumber \\
&\hspace*{0.8in}F_j(f)^\sharp (r,x,v)F_j(f)(r,x+rv_1,v_*)dvdv_*dn dxdr\nonumber \\
&= \int f_0(x,v)dxdv+\mathcal{I}_\epsilon (t)+\int_0^t\int \int _{\mathcal{N}^c _\epsilon} B{\chi}_jff _*F_j(f^\prime )F_j(f^\prime _*)dvdv_*dn dxdr\nonumber \\
&\leq  \int f_0(x,v)dxdv
+\mathcal{I}_\epsilon (t)+\frac{1}{(\gamma \gamma ^\prime \epsilon )^2}\int_0^t\int (n_1^2+n_{\perp 1}^2)[(v-v_*)\cdot n]^2B{\chi}_jff_*\nonumber \\
&\hspace*{0.8in}F_j(f')F_j(f'_*)dvdv_*dn dxdr\nonumber \\
&\leq  \int f_0(x,v)dxdv+\mathcal{I}_\epsilon (t)+\frac{2c_0^\prime }{(\gamma \gamma ^\prime \epsilon )^2}(1+t).
\end{align}
Moreover,
\begin{eqnarray*}
\mathcal{I}_\epsilon (t)\leq 2\pi B_0\epsilon \hspace*{0.03in}t\parallel F_\alpha \parallel _\infty ^2M_j(t)\int f_0(x,v)dxdv.
\end{eqnarray*}
And so, (\ref{dxdv-local}) holds with
\begin{eqnarray*}
c_1^\prime = \max \{ \int f_0(x,v)dxdv, \hspace*{0.02in}\frac{2c_0^\prime }{(\gamma \gamma ^\prime )^2}, \hspace*{0.02in}2\pi B_0\parallel F_\alpha \parallel _\infty ^2\int f_0(x,v)dxdv\} .
\end{eqnarray*}
%
%
\begin{lemma}\label{small-set-local}
\hspace*{0.2in}\\
There is $c_2^\prime $  only depending on $\int f_0(x,v)dxdv$ and $\int |v|^2f_0(x,v)dxdv$ such that, for any $\delta \in ] 0,1[ $,
\begin{equation}\label{control-small-set-local}
\sup _{x_0\in[0,1] }\int_{|x-x_0|<\delta } \hspace*{0.03in}{\sup_{s\in [ 0,t] }}f_j^\sharp (s,x,v)dxdv\leq c_2^\prime \Big( \delta ^{\frac{2}{5}}+t^{\frac{8}{11}}(1+t)^{\frac{3}{11}}\big(1+M_j(t)\big) \Big) ,\quad t>0,\quad j\in \N ^*.
\end{equation}
\end{lemma}
\underline{Proof of Lemma \ref{small-set-local}.} \\
Denote $f_j$ by $f$ for simplicity. For $s\in [0,t] $ it holds,
\[ \begin{aligned}
 f^\sharp (s,x,v)&=f^{\sharp}(t,x,v)-\int_{s}^tQ_j(f)^\sharp (r,x,v)dr\leq f^{\sharp}(t,x,v)+\int_{s}^t(Q_j^-(f))^\sharp (r,x,v)dr,
\end{aligned}\]
where $Q_j^-$ is defined in (\ref{Qj-}). And so
\begin{align}
&\sup_{s\in [ 0,t] }f^\sharp (s,x,v)\leq  f^{\sharp}(t,x,v)\nonumber \\
&+\int_0^t\int B{\chi}_jf^\sharp (r,x,v)f(r,x+rv_1 ,v_*)F_j(f)(r,x+rv_1,v^\prime )F_j(f)(r,x+rv_1,v^\prime _*)dv_*dn dr.\label{Qminus-local}
\end{align}
Denote by
\begin{align*}
\mathcal{J}_\epsilon (t)= \sup _{x_0\in [ 0,1] }\int_0^{t}\int_{|x-x_0|<\delta }\int\int_{\mathcal{N}_\epsilon} &B{\chi}_j
f^\sharp (r,x,v)f(r,x+rv_1 ,v_*)\\
&F_j(f)(r,x+rv_1,v^\prime )F_j(f)(r,x+rv_1,v^\prime _*)dvdv_*dn dxdr.
\end{align*}
Integrating (\ref{Qminus-local}) with respect to $(x,v)$, using Lemma \ref{bony}, the $\frac{1}{\alpha }$ (resp. $\alpha ^{\alpha -1}$) bound from above of $f$ (resp. $F_j(y)$,$y\in [0,\frac{1}{\alpha }]$), gives for any $x_0\in [0,1]$, $\lambda >0$ and $\Lambda >0$ that
\begin{align}\label{lemma4.2-1}
&\int_{|x-x_0|<\delta } \sup_{s\in [ 0,t] }f^\sharp (s,x,v)dxdv\leq \int_{|x-x_0|<\delta } f^{\sharp}(t,x,v)dxdv+\mathcal{J}_\epsilon (t)\nonumber \\
&+\frac{1}{(\lambda \gamma ^{\prime }\epsilon )^2}\int_0^t\int_{|v-v_*|\geq\lambda} (n_1^2+n_{\perp 1}^2)[(v-v_*)\cdot n]^2B{\chi}_jff_*F_j(f')F_j(f'_*)dvdv_*dn dxds\nonumber \\
&+c\int_{0}^{t}\int _{|v-v_*|<\lambda} B{\chi}_j
f^{\sharp}(s,x,v)f(s,x+sv_1,v_*)dvdv_*dn dxds\nonumber \\
\leq  &\int_{|x-x_0|<\delta } f^{\sharp}(t,x,v)dxdv+\mathcal{J}_\epsilon (t)
+\frac{c_0^\prime (1+t)}{(\lambda \gamma ^\prime \epsilon )^2}+ct \lambda^3\int f_0(x,v)dxdv\nonumber \\
\leq &\frac{1}{\Lambda ^2}\int v^2f_0dxdv + c\delta \Lambda^3+\mathcal{J}_\epsilon(t)+\frac{c_0^\prime (1+t)}{(\lambda \gamma ^\prime \epsilon )^2}+ct\lambda^3\int f_0 (x,v)dxdv\nonumber \\
\leq &c\big( \delta ^{\frac{2}{5}}+t^{\frac{2}{5}}\epsilon ^{-\frac{6}{5}}(1+t)^{\frac{3}{5}}\big)  +\mathcal{J}_\epsilon(t),
\end{align}
for an appropriate choice of $(\Lambda ,\lambda )$. Moreover,
\begin{align*}
&\mathcal{J}_\epsilon (t)\leq 2\pi B_0\epsilon t\parallel F_\alpha \parallel _\infty ^2M_j(t)\int f_0(x,v)dxdv.
\end{align*}
Taking $\epsilon = \tilde{c}\big( \frac{1+t}{t}\big) ^{\frac{3}{11}}M^{-\frac{5}{11}}$ with $\tilde{c}$ suitably chosen, leads to
\begin{align*}
&\int_{|x-x_0|<\delta } \sup_{s\in [ 0,t] }f^\sharp (s,x,v)dxdv\leq c\big( \delta  ^{\frac{2}{5}}+t^{\frac{8}{11}}(1+t)^{\frac{3}{11}}M_j(t)^{\frac{6}{11}}\big) .
\end{align*}
The lemma follows.   \cqfd
%
%
\begin{lemma}\label{df-T-local}
\hspace*{0.1in}\\
There is $T>0$ such that the solutions $f_j$ of (\ref{eq-f_j}) satisfy
\begin{eqnarray*}
\int \sup_{(t,x)\in [0,T] \times [ 0,1] }f^\sharp _j(t,x,v)dv\leq 2c_0,\quad j\in \N ^*,
\end{eqnarray*}
with $c_0$ defined in (\ref{Hyp2-f0}).
\end{lemma}
\underline{Proof of Lemma \ref{df-T-local}.} \\
Denote by $E(x)$ the integer part of $x\in\R$, $E(x)\leq x< E(x)+1$. As in (\ref{control-by-gain}),
\begin{align}\label{in1-local}
&\sup_{s\in [ 0,t] }f^\sharp (s,x,v)
\leq  f_0(x,v)\nonumber \\
&+\int_0^t\int B{\chi}_j f(s,x+sv_1,v')f(s,x+sv_1,v'_*)(F_j(f))^\sharp (s,x,v)F_j(f)(s,x+sv_1,v_*)dv_*dnds\nonumber \\
&\leq  f_0(x,v)+\parallel F_\alpha  \parallel _\infty ^2(A_1+A_2+A_3+A_4),
\end{align}
where, for $\epsilon >0$, $\delta  >0$ and $\lambda $ that will be fixed later,
\begin{align*}
&A_1= \int_0^t\int _{|n_1| \geq \epsilon ,\hspace*{0.02in}t|v_1-v^\prime _1|>\delta  }B{\chi}_j\sup _{\tau \in [ 0,t]} f^{\#}(\tau ,x+s(v_1-v^\prime _1),v') \nonumber \\
&\hspace*{2.3in}\sup _{\tau \in [ 0,t]}f^{\#}(\tau ,x+s(v_1-v^\prime _{*1}),v^\prime _*)dv_*dn ds,\\
&A_2= \int_0^t\int _{|n_1| \geq \epsilon ,\hspace*{0.02in}t|v_1-v^\prime _1|<\delta ,\hspace*{0.02in} |v^\prime |<\lambda }B{\chi}_j\sup _{\tau \in [ 0,t]} f^{\#}(\tau ,x+s(v_1-v'_1),v') \times \\
& \hspace*{2.3in}\times \sup _{\tau \in [ 0,t]}f^{\#}(\tau ,x+s(v_1-v^\prime _{*1}),v^\prime _*)dv_*dn ds,\\
&A_3= \int_0^t\int _{|n_1| \geq \epsilon ,\hspace*{0.02in}t|v_1-v^\prime _1|<\delta ,\hspace*{0.02in} |v^\prime |>\lambda }B{\chi}_j\sup _{\tau \in [ 0,t]} f^{\#}(\tau ,x+s(v_1-v'_1),v') \times \\
& \hspace*{2.3in}\times \sup _{\tau \in [ 0,t]}f^{\#}(\tau ,x+s(v_1-v^\prime _{*1}),v^\prime _*)dv_*dn ds,\\
&A_4= \int_0^t\int _{|n_1|<\epsilon }B{\chi}_j\sup _{\tau \in [ 0,t]} f^{\#}(\tau ,x+s(v_1-v'_1),v') \sup _{\tau \in [ 0,t]}f^{\#}(\tau ,x+s(v_1-v^\prime _{*1}),v^\prime _*)dv_*dn ds.
\end{align*}
In $A_1$, $A_2$ and $A_3$, bound the factor $\sup _{\tau \in [ 0,t] }f^\sharp (\tau ,x+s(v_1-v^\prime _{*1}),v^\prime _*)$ by its supremum over $x\in [ 0,1] $, and make the change of variables
\begin{eqnarray*}
 s\rightarrow y= x+s(v_1-v^\prime _1),
\end{eqnarray*}
with Jacobian
\begin{eqnarray*}
\frac{Ds}{Dy}= \frac{1}{|v_1-v^\prime _1|}= \frac{1}{|v-v_*|\hspace*{0.03in}|(n,\frac{v-v_*}{|v-v_*|})|\hspace*{0.03in}|n_1|} \leq \frac{1}{\epsilon \gamma \gamma ^\prime }\hspace*{0.02in}.
\end{eqnarray*}
Consequently,
\begin{align*}
&\sup _{x\in [0,1] }A_1(t,x,v)\\
&\leq \sup _{x\in [0,1] }\int _{t|v_1-v^\prime _1|>\delta }\frac{B{\chi}_j}{|v_1-v^\prime _1|}\Big( \int _{y\in (x,x+t(v_1-v^\prime _1))}\sup _{\tau \in [ 0,t]} f^{\#}(\tau ,y,v')dy\Big)  \nonumber \\
&\hspace*{2.2in}\sup _{(\tau ,X)\in [ 0,t]\times [ 0,1] }f^{\#}(\tau ,X,v'_*)dv_*dn \\
&\leq \int _{t|v_1-v^\prime _1|>\delta }\frac{B{\chi}_j}{|v_1-v^\prime _1|} | E(t(v_1-v^\prime _1)+1)|\Big( \int _0^1\sup _{\tau \in [ 0,t]} f^{\#}(\tau ,y,v^\prime )dy\Big)  \nonumber \\
&\hspace*{2.2in}\sup _{(\tau ,X)\in [ 0,t]\times [ 0,1] }f^{\#}(\tau ,X,v^\prime _*)dv_*dn.
\end{align*}
Performing the change of variables $(v,v_*,n )\rightarrow (v^\prime ,v^\prime _*,-n )$,
\begin{align}
&\int \sup _{x\in [ 0,1] }A_1(t,x,v)dv\nonumber \\
&\leq \int _{t|v_1-v^\prime _1|>\delta }\frac{B{\chi}_j}{|v_1-v^\prime _1|} | E(t(v^\prime _1-v_1)+1)|\Big( \int _0^1\sup _{\tau \in [ 0,t]} f^{\#}(\tau ,y,v)dy\Big)  \nonumber \\
&\hspace*{2.2in}\sup _{(\tau ,X)\in [ 0,t]\times [ 0,1] }f^{\#}(\tau ,X,v_*)dvdv_*dn \nonumber \\
&\leq t(1+\frac{1}{\delta })\int B{\chi}_j\Big( \int _0^1\sup _{\tau \in [ 0,t]} f^{\#}(\tau ,y,v)dy\Big)  \sup _{(\tau ,X)\in [ 0,t]\times [ 0,1] }f^{\#}(\tau ,X,v_*)dvdv_*dn\nonumber \\
&\leq 4\pi B_0\hspace*{0.02in} t(1+\frac{1}{\delta})\Big( \int \sup _{\tau \in [ 0,t]} f^{\#}(\tau ,y,v)dydv\Big) M_j(t).\nonumber
\end{align}
Apply Lemma \ref{integral-dxdv-local}, so that
\begin{equation}\label{bdd-A1-local}
\int \sup _{x\in [ 0,1] }A_1(t,x,v)dv\leq 4\pi B_0c_1^\prime t (1+\frac{1}{\delta })\Big( (1+\frac{1}{\epsilon ^2})(1+t)+\epsilon tM_j(t)\Big) M_j(t) .
\end{equation}
%
%
Moreover,
\begin{align}
c\hspace*{0.01in}\epsilon \int \sup _{x\in [0,1] }A_2(t,x,v)dv&\leq \frac{\delta }{\alpha }\int \int _{|v^\prime |<\lambda }B\chi _j\sup _{(\tau ,X)\in [ 0,t] \times [ 0,1] } f^{\#}(\tau ,X,v^\prime _*)dvdv_*dn\nonumber \\
&= \frac{\delta }{\alpha }\int _{|v|<\lambda }\int B\chi _j\sup _{(\tau ,X)\in [ 0,t] \times [ 0,1] } f^{\#}(\tau ,X,v_*)dvdv_*dn\nonumber \\
& \hspace*{1.in}\text{by the change of variables }(v,v_*,n)\rightarrow (v^\prime ,v^\prime _*,-n)\nonumber \\
&\leq \frac{c\delta \lambda ^3}{\alpha }M_j(t), \label{A2-local}
\end{align}
and
\begin{align}
&c\hspace*{0.01in}\epsilon \int \sup _{x\in [0,1] }A_3(t,x,v)dv\nonumber \\
&\leq \int _{|v^\prime |>\lambda }B\chi _j\Big( \int _0^1\sup _{\tau \in [ 0,t] } f^{\#}(\tau ,y,v^\prime )dy\Big) \sup _{(\tau ,X)\in [ 0,t] \times [ 0,1] } f^{\#}(\tau ,X,v^\prime _*)dvdv_*dn\nonumber \\
&\leq c\Big( \int _0^1\int _{|v|>\lambda }\sup _{\tau \in [ 0,t] } f^{\#}(\tau ,y,v)dvdy\Big) \int \sup _{(\tau ,X)\in [ 0,t] \times [ 0,1] } f^{\#}(\tau ,X,v_*)dv_*\nonumber \\
&\hspace*{1.in}\text{by the change of variables }(v,v_*,n)\rightarrow (v^\prime ,v^\prime _*,-n)\nonumber \\
&\leq \frac{c}{\lambda ^2}M_j^2(t)\quad \text{by Lemma \ref{mass-tails}}. \label{A3-local}
\end{align}
Finally, with the change of variables $(v,v_*,n)\rightarrow (v^\prime ,v^\prime _*,-n)$,
\begin{align}
\int \sup _{x\in [0,1] }A_4(t,x,v)dv&\leq B_0t\big( \int _{|n_1| <\epsilon }dn\big) \Big( \int \sup _{(\tau ,x)\in [ 0,t] \times [ 0,1] } f^{\#}(\tau ,x,v)dv\Big) ^2\nonumber \\
& \leq 2\pi B_0\epsilon \hspace*{0.02in}t M_j^2(t).\label{A4-local}
\end{align}
It follows from (\ref{in1-local}), (\ref{bdd-A1-local}), (\ref{A2-local}), (\ref{A3-local}) and (\ref{A4-local}) that
\hspace*{0.1in}\\
\begin{equation}\label{trinome-M}
a(t)M^2_j(t)-b(t)M_j(t)+c_0\geq 0,\quad t\leq 1,
\end{equation}
where for some positive constants $(c^\prime _l)_{2\leq l\leq 4}$ independent on $\epsilon $, $\delta $ and $\lambda $,
\begin{align*}
&a(t)= c^\prime _2\big( \epsilon t(1+\delta ^{-1})+\epsilon ^{-1}\lambda ^{-2}\big) ,\quad b(t)= 1-c^\prime _3t(1+\delta ^{-1})(1+\epsilon ^{-2})-c^\prime _4\epsilon ^{-1}\delta \lambda ^3.
\end{align*}
Choose $\lambda = \epsilon ^{-1}$, $\delta = \epsilon ^5$ and $\epsilon = \frac{1}{16}\min \{ \frac{1}{c_4^\prime },\frac{1}{c_0}\} $. For $T$ small enough, it holds that
\begin{equation}\label{sufficient-local}
b(t)\in \hspace*{0.03in}] \frac{3}{4}, 1[ \quad \text{and   }c_0a(t)<\frac{1}{8},\quad t\in [0,T] ,
\end{equation}
which is sufficient  for the polynomial in (\ref{trinome-M}) to have two nonnegative roots and take a negative value at $2c_0$. Recalling that  $M_j(0)= c_0$ and $M_j$ is continuous by the continuity in time and space of $f_j$, it follows that
\begin{eqnarray*}
M_j(t)\leq 2c_0, \quad t\in [0,T] .
\end{eqnarray*}
\cqfd
%
%
%
%
%
\section{Control of the mass density under the assumptions of Theorem \ref{th-global}.}\label{mass-density-global}
\setcounter{theorem}{0}
\setcounter{equation}{0}
Let $k\in \{ 1, 2, 3\} $.
Under the supplementary assumption (\ref{hyp4-B}), we prove a uniform control with respect to $j$ of the mass density $M_j(t)$ defined in (\ref{df-mass-density}). It relies on the two following lemmas.
%
%
\begin{lemma}\label{integral-dxdv-global}
\hspace*{0.2in}\\
Given $\epsilon>0$, there exists a constant $c'_5$ only depending on
$\int f_0(x,v)dxdv$, such that
\begin{equation}\label{dxdv}
\int \sup_{s\in [0,t]}f_j^\sharp (s,x,v)dxdv\leq c'_5(1+t) ,\quad t>0,\quad j\in \N ^*.
\end{equation}
\end{lemma}
\underline{Proof of Lemma \ref{integral-dxdv-global}}\\
Denote $f_j$ by $f$ for simplicity. By the non-negativity of $f$, it holds
\begin{align*}
f^\#(s, x, v)\leq &  f_0(x,v)+\int_0^s\int f^\sharp (\tau ,x+\tau (\bar{v}-\overline{v^\prime }),v^\prime ) f^\sharp (\tau ,x+\tau (\bar{v}-\overline{v^\prime _*}),v^\prime _*)\times \\
&\times F_j(f^\sharp (\tau ,x,v))F_j(f(\tau ,x+\tau (\bar{v}-\overline{v_*}), v_*))B_1(v-v_*)B_2(\theta)dv_*dnd\tau .\nonumber
\end{align*}
Using the $\frac{1}{\alpha }$ bound for $f^\sharp (\tau ,x+\tau (\bar{v}-\overline{v^\prime _*}),v^\prime _*)$, and (\ref{hyp4-B})  leads to
\begin{align}
&\sup_{s\in [0,t]} f^\#(s, x, v)\leq f_0(x,v)+c\int_0^t \int f^\#(s,x+s(\bar{v}-\overline{v^\prime }),v^\prime )B_1(v-v_*)B_2(\theta)dv_*dnds. \label{lemma5.1-1}
\end{align}
Hence,
\begin{align*}
&\int \sup_{s\in [0,t]} f^\#(t,x,v) dxdv\\
&\leq\int f_0(x,v)dxdv +c\int_0^t  \int f^\#(s,x+s(\bar{v}-\overline{v^\prime }),v^\prime )B_1(v-v_*)B_2(\theta)dxdv_*dvdnds\\
&=\int f_0(x,v)dxdv +c\int_0^t \int f(s,x,v)B_1(v-v_*)B_2(\theta)dxdv_*dvdnds\\
&\leq\int f_0(x,v)dxdv
 +c\int_0^t\int_\gamma^\infty\int f(s,x,v)r^{-(1+\eta)}dxdvdrds\quad \text{by  }(\ref{hyp4-B}) \\
 &= \int f_0(x,v)dxdv+\frac{c}{\eta \gamma^\eta}\int_0^t\int  f(s,x,v)dxdvds\\
 &=: c^\prime _5(1+t),
\end{align*}
by the mass conservation.
\cqfd
%
%
\begin{lemma}\label{df-T-global}
\hspace*{0.1in}\\
Given $T>0$, the solutions $f_j$ of (\ref{eq-f_j}) satisfy
\begin{eqnarray*}
M_j(T)\leq c_1(T),\quad j\in \N ^*,
\end{eqnarray*}
where $c_1(T)$ only depends on $T$ and $c_0$.
\end{lemma}
{\underline{Proof of Lemma \ref{df-T-global}.}\\
By (\ref{lemma5.1-1}), for any $(t,x)\in [0,T] \times \R ^3$,
\begin{eqnarray*}
\sup_{(s,x)\in [0,t]\times [0,1]^k} f^\#(s,x,v)\leq \sup_{x\in [0,1]^k}f_0(x,v)+c\int_0^t \int\sup_{x\in [0,1]^k} f(s,x,v^{\prime })B_1(v-v_*)B_2(\theta)dv_*dnds.
\end{eqnarray*}
Consequently,
\begin{align*}
\int \sup_{(s,x)\in [0,t] \times [0,1]^k} f(s,x,v) dv&\leq c_0+c\int_0^t\int \sup_{x\in [0,1]^k} f(s,x,v^\prime )B_1(v-v_*)B_2(\theta)dv_*dvdnds\\
&= c_0+c\int_0^t\int \sup_{x\in [0,1]^k} f(s,x,v)B_1(v-v_*)B_2(\theta)dv_*dvdnds\\
&\leq  c_0+\frac{c}{\eta \gamma^\eta}\int_0^t\int\sup_{x\in [0,1] ^k} f(s,x,v)dvds.
\end{align*}
It follows that
\begin{eqnarray*}
\int \sup_{(t,x)\in [0,T]\times [0,1]^k}f(t,x,v) dv\leq c_0e^{c^{\prime \prime }T},\quad \text{with   }c^{\prime \prime }= \frac{c}{\eta \gamma ^\eta }.
\end{eqnarray*}
\cqfd
}
%
%
%
\section{Well-posedness of the Cauchy problem.}\label{pf-th}
\setcounter{theorem}{0}
\setcounter{equation}{0}
Let $T_0$ be supremum of the times up to which it has been proved that the mass densities of the approximations are uniformly bounded. Recall that $T_0$ may be finite (resp. is infinite) under the assumptions of Theorem \ref{th-local} (resp. \ref{th-global}). We prove in this section that for any $T\in [0, T_0[$ there is a unique solution to the Cauchy problem (\ref{f})-(\ref{init}).
This section is divided into three steps. In the first step, we study initial layers for the approximations. In the second step, the existence of a solution $f$ to (\ref{f}) on $[ 0,T] $ for $T\in ] 0,T_0[ $ is shown. Finally the third step proves the uniqueness and stability result stated in Theorems \ref{th-local} and \ref{th-global}.\\
\hspace*{0.1in}\\
\underline{First step: initial layers.}
%
%
\begin{lemma}\label{initial-layer}
\hspace*{0.1in}\\
For any $T\in [0, T_0[$, there are $j_T\in \N ^*$, a positive time $t_m>0$, and for $V>0$ positive constants $b_V$ and $\mu _V$ such that
\begin{align*}
& f_j^\sharp (t,\cdot ,v)\leq \frac{1}{\alpha }-b_V\hspace*{0.01in}t,\quad t\in [ 0,t_m] ,\quad \lvert v\rvert <V,\quad j\geq j_T,\\
& f_j^\sharp (t,\cdot ,v)\leq \frac{1}{\alpha }-\mu _V,\quad t\in [t_m,T] ,\quad \lvert v\rvert <V,\quad j\geq j_T.
\end{align*}
\end{lemma}
\underline{Proof of Lemma \ref{initial-layer}.} \\
Denote $f_j$ by $f$ for simplicity. It follows
from Lemmas \ref{df-T-local} and \ref{df-T-global} that there is $c_1(T)>0$ such that 
\begin{equation}\label{lemma6.1-1}
M_j(T)\leq c_1(T),\quad j\in \N ^*.
\end{equation}
Denote by
\begin{eqnarray*}
\tilde{\nu}_j(f):= \int B\chi _j f^\prime f^\prime _*F_j(f_*) dv_*dn ,\quad \nu _j( f):= \int B\chi _j f_*F_j(f^\prime )F_j(f^\prime _*) dv_*dn ,
\end{eqnarray*}
so that
\begin{eqnarray*}
Q_j(f)= F_j(f)\tilde{\nu }_j(f)-f\nu _j(f) .
\end{eqnarray*}
It follows from (\ref{lemma6.1-1}) that  $\nu _j(f)^\sharp $ and $\tilde{\nu}_j(f)^\sharp $ are bounded from above uniformly with respect to $j$. Denote by $c_2(T)$ a bound from above of ($\tilde{\nu}_j(f)^\sharp )_{j\in \N }$.\\
Let us prove that $(\nu _j(f)^\sharp )$ is bounded from below for large $j$ on $[0,T] \times [0,1]^k \times \{ v; \lvert v\rvert <V\} $ for any $V>0$. By definition,
\[ \begin{aligned}
&\nu _j(f)^\sharp (t,x,v)= \int B\chi_j f(t,x+t\bar{v},v_*)F_j(f(t,x+t\bar{v},v^\prime ))F_j(f(t,x+t\bar{v},v^\prime _*))dv_*dn .
\end{aligned}\]
Using Duhamel's form for the solution, (\ref{lemma6.1-1}) and (\ref{Hyp4-f0}), one gets that
\begin{equation}\label{bdd-below-fj1}
f(t,x+t\bar{v},v_*)\geq c_3(T) f_0(x,v_*)>0,\quad \text{a.a. }(t,x,v,v_*)\in [0,T]\times [0,1] ^k\times \R ^3\times \R ^3,
\end{equation}
for some constant $c_3(T)>0$. For any angles $(\theta , \varphi )\in [0, 2\pi ] \times [0, \pi ]$ defining the relative position of $v^\prime -v$ with respect to $v_*-v$, the maps $v_*  \rightarrow v^\prime $ and $v_*  \rightarrow v^\prime _*$ are changes of variables. Indeed, consider the map $v_*  \rightarrow v^\prime $, reduce it to $v_* -v \rightarrow v^\prime -v$ and denote it by $U$ . Let $n$ be the vector with polar coordinates $(\theta , \varphi )$ with respect to $v_*-v$. Choose a coordinates system with the first (resp. second, resp. third) axis in the direction of $v_*-v$ (resp. orthogonal to $v_*-v$ in the plane defined by $v_*-v$ and $n$, resp. orthogonal to the two first axes). The map $U$ maps the volume $d(v_{*x}-v_x)d(v_{*y}-v_y)d(v_{*z}-v_z)$ into
\begin{align*}
d(v^\prime _{x}-v_x)d(v^\prime _{y}-v_y)d(v^\prime _{z}-v_z)&= (\cos \theta )^4d(v_{*x}-v_x)d(v_{*y}-v_y)d(v_{*z}-v_z)\\
&+O\Big( \big( d(v_{*x}-v_x)\big) ^2+\big( d(v_{*y}-v_y)\big) ^2+\big( d(v_{*z}-v_z)\big) ^2\Big) ,
\end{align*}
since up to second order terms with respect to $d(v_{*x}-v_x)$, $d(v_{*y}-v_y)$ and $d(v_{*z}-v_z)$, the length $d(v_{*x}-v_x)$ (resp. $d(v_{*y}-v_y)$, resp. $d(v_{*z}-v_z)$) is changed into $|\cos \theta |d(v_{*x}-v_x)$ \\
(resp. $|\cos \theta |d(v_{*y}-v_y)$,
resp. $\cos ^2\theta d(v_{*z}-v_z)$). And so the Jacobian of $U$ equals $\cos ^4\theta $. Using these changes of variables and
%
%
 (\ref{lemma6.1-1}), it holds that
 \begin{align*}
 \int f(t,x+t\bar{v},v^\prime )dv_*<&\frac{c_1(T)}{(\gamma ^\prime )^4}\text{  and } \int f(t,x+t\bar{v},v^\prime _*)dv_*<\frac{c_1(T)}{(\gamma ^\prime )^4},\\
 & \text{a.a.  }(t,x,v,\theta ,\varphi )\in [0,T] \times [0,1] ^k\times \R ^3\times [0, 2\pi ] \times [0, \pi ] ,\quad |\cos \theta | >\gamma ^\prime .
 \end{align*}
Consequently, the measure of the set
\begin{equation}\label{df-Z-j0}
Z_{(j,t,x,v,\theta ,\varphi )}:= \{v_*;f(t,x+t\bar{v},v')>\frac{1}{2}\quad \text{or}\quad f(t,{x}+t\bar{v},v'_*)>\frac{1}{2}\}
\end{equation}
is bounded by $\frac{2c_1(T)}{(\gamma ^\prime )^4}$, uniformly with respect to $(x,v,\theta ,\varphi )$ with $|\cos \theta | >\gamma ^\prime $, $t\in [0,T]$, and $j\in \N ^*$. Take $j_T$ so large that $\frac{4}{3}\pi j_T^3$ is at least twice this uniform bound.  Notice that here  $j_T$ only depends on $T$, $\int f_0(x,v)dxdv$ and $\int |v|^2f_0(x,v)dxdv$. Denote by $\mathcal{B}\big( 0,\big( \frac{3c_1(T)}{\pi (\gamma ^\prime )^4}\big) ^{\frac{1}{3}}\big) $ the ball of radius $\big( \frac{3c_1(T)}{\pi (\gamma ^\prime )^4}\big) ^{\frac{1}{3}}$. 
It follows from (\ref{bdd-below-fj1}) and the definition of $j_T$ that
\begin{align*}
&\nu _j(f)^\sharp (t,x,v)\nonumber \\
&\geq \int _{\mathcal{S}^2}\int _{\mathcal{B}\big( 0,\big( \frac{3c_1(T)}{\pi (\gamma ^\prime )^4}\big) ^{\frac{1}{3}}\big) \cap Z^c_{(j,t,x,v,\theta ,\varphi )}}B\chi_j f(t,x+t\bar{v},v_*)F_j(f(t,x+t\bar{v},v^\prime ))\nonumber \\
&\hspace*{3.4in}F_j(f(t,x+t\bar{v},v^\prime _*))dv_*dn\nonumber \\
&\geq c_3(T)(1-\frac{\alpha }{2})^{2\alpha }\int _{\mathcal{S}^2}\int _{\mathcal{B}\big( 0,\big( \frac{3c_1(T)}{\pi (\gamma ^\prime )^4}\big) ^{\frac{1}{3}}\big) \cap Z^c_{(j,t,x,v,\theta ,\varphi )}}B(\lvert v-v_*\rvert ,\theta )\inf _{x\in [0,1]^k}f_0(x,v_*)dv_*dn,\nonumber \\
&\hspace*{2.in}j\geq j_T,\quad \text{a.a.  }(t,x,v)\in [ 0,T] \times [ 0,1]^k \times \{ v\in \R ^3;|v|<V\} .
\end{align*}
Using a median property for the restriction of $v\rightarrow \inf _{x\in [0,1]^k}f_0(x,v)$ to the ball $\mathcal{B}\big( 0,\big( \frac{3c_1(T)}{\pi (\gamma ^\prime )^4}\big) ^{\frac{1}{3}}\big) $, which is a bounded measurable Lebesgue function, there are two disjoint sets $\Omega _1$ and $\Omega _2$ of equal volume, such
that 
\begin{align*}
&\inf _{x\in [0,1]^k}f_0(x,v_1)\leq \inf _{x\in [0,1]^k}f_0(x,v_2) \text{  for a.a. }v_1\in \Omega _1,\quad v_2\in \Omega_2.
\end{align*}
Denote by $\Gamma = V+(\frac{3c_1(T)}{\pi (\gamma ^\prime )^4})^{\frac{1}{3}}$. \\
For $j\geq j_T$ and a.a. $(n,t,x,v)\in \mathcal{S}^2\times [ 0,T] \times [ 0,1]^k \times \{ v\in \R ^3;|v|<V\} $,
\begin{align*}
&\int _{\mathcal{B}\big( 0,\big( \frac{3c_1(T)}{\pi (\gamma ^\prime )^4}\big) ^{\frac{1}{3}}\big) \cap Z^c_{(j,t,x,v,\theta ,\varphi )}}B(\lvert v-v_*\rvert ,\theta )\inf _{x\in [0,1]^k}f_0(x,v_*)dv_*\nonumber \\
&\geq \inf _{u\in [\gamma ,\Gamma ]}B(u,\theta )\inf _{\overline{\Omega }\subset \mathcal{B}\big( 0,\big( \frac{3c_1(T)}{\pi (\gamma ^\prime )^4}\big) ^{\frac{1}{3}}\big) ; \hspace*{0.02in}\lvert \overline{\Omega }\rvert = \frac{2c_1(T)}{(\gamma ^\prime )^4}}\hspace*{0.04in}\int _{\overline{\Omega }}\inf _{x\in [0,1]^k}f_0(x,v_*)dv_*\nonumber \\
&= \inf _{u\in [\gamma ,\Gamma ]}B(u,\theta )\hspace*{0.03in}\int _{\Omega _1}\inf _{x\in [0,1]^k}f_0(x,v_*)dv_*.
\end{align*}
Hence, by (\ref{hyp3-B}), for $j\geq j_T$ and a.a. $(t,x,v)\in [ 0,T] \times [ 0,1]^k \times \{ v\in \R ^3;|v|<V\} $,
\begin{align}\label{bdd-below-nu}
\nu _j(f)^\sharp (t,x,v)&\geq c_3(T)(1-\frac{\alpha }{2})^{2\alpha }\Big( \int _{\mathcal{S}^2}\inf _{u\in [\gamma ,\Gamma ]}B(u,\theta )dn\Big) \int _{\Omega _1 }\inf _{x\in [0,1]^k}f_0(x,v_*)dv_*\nonumber \\
&\geq c_\Gamma c_3(T)(1-\frac{\alpha }{2})^{2\alpha }\int _{\Omega _1 }\inf _{x\in [0,1]^k}f_0(x,v_*)dv_*.
\end{align}
Applying (\ref{Hyp4-f0}) to $\Omega _1$, this is a positive bound from below of $\big( \nu _j(f)^\sharp (t,x,v)\big) _{j\geq j_T}$ \\
on $[ 0,T] \times [ 0,1]^k \times \{ v\in \R ^3;|v|<V\} $.\\
The functions defined on $] 0,\frac{1}{\alpha }] $ by $x\rightarrow \frac{F_j(x)}{x}$ are uniformly bounded {from above} with respect to $j$ by
\begin{eqnarray*}
x\rightarrow \alpha ^{\alpha -1}\frac{(1-\alpha x)^\alpha }{x}, 
\end{eqnarray*}
that is continuous and decreasing to zero at $x= \frac{1}{\alpha }$. Hence there is $\tilde{\mu }_V= \min \{ \frac{1}{2\alpha }, (\frac{c_4(T)c_\Gamma }{2c_2(T)})^{\frac{1}{\alpha }}\}$ such that
\begin{eqnarray*}
x\in{ [ } \frac{1}{\alpha }-\tilde{\mu }_V,\frac{1}{\alpha }] \quad \text{implies}\quad \frac{F_j(x)}{x}{\leq \frac{c_4(T)c_\Gamma }{4c_2(T)}},\quad j\geq j_T.
\end{eqnarray*}
Consequently, for $j\geq j_T$ and $\lvert v\rvert <V$,
\begin{align}
f^\sharp (t,x,v)\in { [ } \frac{1}{\alpha }-\tilde{\mu }_V,\frac{1}{\alpha }] \quad \Rightarrow \quad D_tf^\sharp (t,x,v)&= \big( F_j(f^\sharp )\tilde{\nu }_j^\sharp -\frac{1}{2}f^\sharp \nu _j^\sharp \big) (t,x,v)-\frac{1}{2}f^\sharp \nu _j^\sharp (t,x,v)\nonumber \\
&< -\frac{1}{2}f^\sharp \nu _j^\sharp (t,x,v)\hspace*{1.8in}\nonumber \\
&<-\frac{c_4(T)c_\Gamma }{4\alpha }:=-b_V.
\end{align}
This gives a maximum time $t_1=\frac{{\tilde{\mu }_V}}{b}$ for $f^\#$ to reach $\frac{1}{\alpha}-\tilde{\mu }_V$ from an initial value $f_0(x,v)\in ] \frac{1}{\alpha }-\tilde{\mu }_V,\frac{1}{\alpha }]$.  On this time interval $D_tf^\sharp \leq -b_V$. If $t_1\geq T$, then at $t= T$ the value of $f^\#$ is bounded from above by $\frac{1}{\alpha}-b_VT:= \frac{1}{\alpha}-\mu ^\prime _V$ with $0<\mu^\prime \leq \tilde{\mu }_V$. Let
\begin{eqnarray*}
t_m=\min \{ t_1,T\} , \quad \mu _V= \min \{ \tilde{\mu }_V, \mu ^\prime _V\} .
\end{eqnarray*}
For any  $(x,v)$ with $\lvert v\rvert <V$, if $f(0,x,v)< \frac{1}{\alpha }-\mu _V$ were to reach $\frac{1}{\alpha}-\mu _V$ at $(t,x,v)$ with $t\leq t_m$, then 
$D_tf^\#(t,x,v)\leq -b_V$, which excludes such a possibility.
It follows that 
\begin{align}\label{interval0-tm}
&f^\sharp (t,x,v)\leq \frac{1}{\alpha}-\mu  _V\text{     for     } j\geq j_T, (t,x)\in [ t_m, T]\times [0,1]^k,\quad  \lvert v\rvert <V,\nonumber \\
&f^\sharp (t,x,v)\leq \frac{1}{\alpha }-b_V\hspace*{0.01in}t\quad \text{for   }j\geq j_T, (t,x)\in [0,t_m]\times [0,1]^k, \lvert v\rvert <V.
\end{align}
The previous estimates leading to the definition of $t_m$ are independent of $j\geq j_T$.\\
\cqfd
\underline{Second step: existence of a solution $f$ to (\ref{f}).}
\hspace*{0.1in}\\
\hspace*{0.1in}\\
Let $T\in [0,T_0[ $ where $T_0$, defined at the beginning of this section, may be finite under the hypothesis of Theorem 2.1 and is infinite under those of Theorem 2.2. We shall prove the convergence in $L^1([ 0,T] \times [ 0,1]^k \times \R ^3))$ of the sequence $(f_j)$ to a solution $f$ of (\ref{f}) by proving that  it is a Cauchy sequence. Let us first prove that it is a Cauchy sequence in $L^1([ 0,T_0] \times [ 0,1]^k \times \R ^3))$ for some $T_0\in ] 0,T[ $, i.e. for any $\beta >0$, there exists $a\geq \max \{ 1, j_T\} $ such that
\begin{equation}
\sup_{{t\in }[0,T_0]}\int |g_j(t,x,v)|dxdv< \beta,\quad j>a,
\end{equation}
where $g_j=f_j-f_a$. The sequence $(f_j)$ will be proven to be a Cauchy sequence in \\
$L^1([ T_0,2T_0] \times [ 0,1]^k \times \R ^3))$ etc. in an analogous way. \\
By the uniform boundedness of energy of $(g_j)$, there is $V>0$ such that
\begin{equation}
\sup_{{t\in }[0,T]}\int _{\lvert v\rvert \geq V}|g_j(t,x,v)|dxdv< \frac{\beta }{2},\quad j>a,
\end{equation}
The function $g_j$ satisfies the equation
\begin{align}\label{cauchy1}
&\partial _tg_j+\bar{v}\cdot \nabla _xg_j\nonumber \\
&= \int ({\chi} _j-{\chi}_a)B\Big( f_j^\prime f^\prime _{j*}F_j(f_j)F_j(f_{j*})- f_jf_{j*}F_j(f^\prime _j)F_j(f^\prime _{j*})\Big) dv_*dn\nonumber \\
&+\int \chi _a B(f_j^\prime f_{j*}^\prime -f_a^\prime f_{a*}^\prime )F_j(f_j)F_j(f_{j*})dv_*dn\nonumber \\
&-\int \chi _a B(f_jf_{j*} -f_af_{a*})F_j(f_j^\prime )F_j(f_{j*}^\prime )dv_*dn\nonumber \\
&+\int \chi _aBf_a^\prime f_{a*}^\prime \Big( F_j(f_{j*})\big( F_j(f_j)-F_j(f_a)\big) +F_a(f_a)\big( F_j(f_{j*})-F_j(f_{a*})\big) \Big) dv_*dn\nonumber \\
&+\int \chi _aBf_a^\prime f_{a*}^\prime \Big( F_j(f_{j*})\big( F_j(f_a)-F_a(f_a)\big) +F_a(f_a)\big( F_j(f_{a*})-F_a(f_{a*})\big) \Big) dv_*dn\nonumber \\
& -\int \chi _aBf_af_{a*}\Big( F_j(f_{j*}^\prime )\big( F_j(f_j^\prime )-F_j(f_a^\prime )\big) +F_a(f_a^\prime )\big( F_j(f_{j*}^\prime )-F_j(f_{a*}^\prime )\big) \Big) dv_*dn\nonumber \\
&-\int \chi _aBf_af_{a*}\Big( F_j(f_{j*}^\prime )\big( F_j(f_a^\prime )-F_a(f_a^\prime )\big) +F_a(f_a^\prime )\big( F_j(f_{a*}^\prime )-F_a(f_{a*}^\prime )\big) \Big) dv_*dn.\quad
\end{align}
Using Lemmas \ref{df-T-local} and \ref{df-T-global} and the conservation of energy of $f_j$,
\begin{align*}
\int ({\chi} _j-{\chi}_a)B&\Big( f_j^\prime f^\prime _{j*}F_j(f_j)F_j(f_{j*})+f_jf_{j*}F_j(f^\prime _j)F_j(f^\prime _{j*})\Big) dxdvdv_*dn\\
& \leq c\int _{\lvert v\rvert >\frac{a}{\sqrt{2}}}f_j(t,x,v)dxdv\\
&\leq \frac{c}{a^2}.
\end{align*}
Moreover,
\begin{align}
&\int \chi _a B\lvert f_jf_{j*} -f_af_{a*}\rvert F_j(f_j^\prime )F_j(f_{j*}^\prime )dxdvdv_*dn\nonumber \\
&\leq c\Big( \int \sup _{(t,x)\in [ 0,T] \times [ 0,1]^k }f_j^\sharp (t,x,v)dv
+ \int \sup _{(t,x)\in [ 0,T] \times [ 0,1]^k }f_a^\sharp (t,x,v)dv\Big) \nonumber \\
&\hspace*{2.45in}\times \Big( \int \lvert (f_j^\sharp -f_a^\sharp )(t,x,v)\rvert dxdv\Big) \nonumber \\
&\leq c\int \lvert (f_j^\sharp -f_a^\sharp )(t,x,v)\rvert dxdv,\quad \text{   by Lemmas \ref{df-T-local} and \ref{df-T-global},}
\end{align}
and
\[ \begin{aligned}
&\int \chi _aB\Big( f_a^\prime f_{a*}^\prime F_j(f_{j*})\lvert  F_j(f_a)-F_a(f_a)\rvert \Big) ^\sharp dxdvdv_*dn = \int \chi _aBf_a^\prime f_{a*}^\prime F_j(f_{j*})\\
&(1-\alpha f_a) (1+(1-\alpha)f_a)^{1-\alpha} \lvert (\frac{1}{j}+1-\alpha f_a)^{\alpha-1}
-(\frac{1}{a}+1-\alpha f_a)^{\alpha-1}\rvert dxdvdv_*dn.
\end{aligned}\]
By Lemmas \ref{integral-dxdv-local}, \ref{df-T-local} and \ref{integral-dxdv-global}, \ref{df-T-global}, this integral restricted to the set where $1-\alpha f_a(t,x,v))\leq\frac{2}{a}$, {hence where
\begin{eqnarray*}
(1-\alpha f_a)  \lvert (\frac{1}{j}+1-\alpha f_a)^{\alpha-1}
-(\frac{1}{a}+1-\alpha f_a)^{\alpha-1}\rvert
\leq \frac{2^{\alpha +1}}{a^\alpha} ,
\end{eqnarray*}
is bounded by $\frac{c}{a^\alpha}$ for some constant $c>0$. } \\
For the remaining domain of integration where $1-\alpha f_a(t,x,v))\geq\frac{2}{a}$, {it holds
\[ \begin{aligned}
|F_j(f_a)-F_a(f_a)| &\leq c (1-\alpha f_a)^\alpha\lvert (\frac{1}{j(1-\alpha f_a)}+1)^{\alpha-1}
-(\frac{1}{a(1-\alpha f_a)}+1)^{\alpha-1}\rvert \\
&= c(\frac{1}{j}-\frac{1}{a})(1-\alpha f_a)^{\alpha -1}\lambda ^{\alpha -2}\quad \text{where    }\lambda \in [ 1,\frac{3}{2}]\\
&\leq \frac{2^{\alpha -1}c}{a^\alpha}.
\end{aligned}\]
And so,
\[ \begin{aligned}
&\int \chi _aB\Big( f_a^\prime f_{a*}^\prime F_j(f_{j*})\lvert  F_j(f_a)-F_a(f_a)\rvert \Big) ^\sharp dxdvdv_*dn \leq \frac{c}{a^\alpha }.
\end{aligned}\]
}
Finally
\begin{align*}
&\int _{\lvert v\rvert <V}\chi _aB\Big( f_a^\prime f_{a*}^\prime F_j(f_{j*})\lvert F_j(f_j)-F_j(f_a)\rvert \Big) ^\sharp (t,x,v)dxdvdv_*dn\\
&\leq  c\int _{\lvert v\rvert <V}\lvert F_j(f_j)-F_j(f_a)\rvert ^\sharp (t,x,v)dxdv+c\beta .
\end{align*}
Split the $(x,v)$-domain of integration of the latest integral into
\[ \begin{aligned}
&D_1:= \{ (x,v); \lvert v\rvert <V \text{ and }(f_j^\sharp (t,x,v),f_a^\sharp (t,x,v))\in [ 0,\frac{1}{\alpha }-\mu _V] ^2\} ,\\
&D_2:= \{ (x,v); \lvert v\rvert <V \text{ and }(f_j^\sharp (t,x,v),f_a^\sharp (t,x,v))\in [ \frac{1}{\alpha }-\mu _V ,\frac{1}{\alpha }] ^2\} ,\\
&D_3:= \{ (x,v); \lvert v\rvert <V,(f_j^\sharp ,f_a^\sharp )(t,x,v)\in [ \frac{1}{\alpha }-\mu _V,\frac{1}{\alpha }] \times [ 0,\frac{1}{\alpha }-\mu _V] \\
&\hspace*{1.25in}\text{   or   } (f_j^\sharp ,f_a^\sharp )\in [ 0,\frac{1}{\alpha }-\mu _V ] \times [ \frac{1}{\alpha }-\mu _V,\frac{1}{\alpha } ]\} .
\end{aligned}\]
It holds that
\[ \begin{aligned}
&\int _{D_1}\lvert F_j(f_j)-F_j(f_a)\rvert ^\sharp (t,x,v)dxdv\leq c(\alpha \mu _V)^{\alpha -1}\int _{D_1}\lvert g_j^\sharp (t,x,v)\rvert dxdv,\\
&\int _{D_2}\lvert F_j(f_j)-F_j(f_a)\rvert ^\sharp (t,x,v)dxdv\leq c(b_Vt)^{\alpha -1}\int _{D_2}\lvert g_j^\sharp (t,x,v)\rvert dxdv,\quad {\text{by   }(\ref{interval0-tm}),}\\
&\int _{D_3}\lvert F_j(f_j)-F_j(f_a)\rvert ^\sharp (t,x,v)dxdv\leq c\big( (\alpha \mu _V)^{\alpha -1}+(b_Vt)^{\alpha -1}\big) \int _{D_3}\lvert g_j^\sharp (t,x,v)\rvert dxdv.
\end{aligned}\]
The remaining terms to the right in (\ref{cauchy1}) are of the same types as the ones just estimated. \\
Consequently,
\begin{equation}\label{ineq-g_j-1}
\frac{d}{dt}\int _{\lvert v\rvert <V}|g_j^\sharp (t,x,v)|dxdv\leq \frac{c}{a^{\alpha }}+c\beta +c(1+\mu _V^{\alpha -1}+(b_Vt)^{\alpha -1})\Big( \int _{\lvert v\rvert <V}\lvert g_j^\sharp (t,x,v)\rvert dxdv\Big) .
\end{equation}
And so,
\begin{align}\label{ineq-g_j-2}
&\sup _{t\in [0,T_0] }\int _{\lvert v\rvert <V}|g_j^\sharp (t,x,v)|dxdv\nonumber \\
&\leq \Big( \int _{\lvert v\rvert <V}\mid (f_{0,j}-f_{0,a})(x,v)\mid dxdv+\frac{cT}{a^{\alpha }}+c\beta T_0\Big) e^{c((1+\mu _V^{\alpha -1})T+\frac{b_V^{\alpha -1}T^\alpha }{\alpha })},
\end{align}
with $f_{0,j}$ (resp. $f_{0,a}$) defined in (\ref{df-init-approx}). For $a$ (resp. $T_0$) large (resp. small) enough, the right-hand side of (\ref{ineq-g_j-2}) is smaller than $\frac{\beta }{2}$, uniformly w.r.t. $j\geq a$. 
{This proves that $(f_j)_{j\in \N ^*}$ is a Cauchy sequence in $L^1([ 0,T_0] \times [ 0,1]^k \times \R ^3)$ and } ends the proof of the existence of a solution $f$ to (\ref{f}). It follows from the boundedness of $\frac{d}{dt}f^\sharp $ that $f\in C([0,T]; L^1([0,1]^k \times \R ^3))$, which in turn implies that $Q(f)\in C([0,T]; L^1([0,1]^k \times \R ^3))$ and $f\in C^1([0,T]; L^1([0,1]^k \times \R ^3))$.\\
\hspace*{0.1in}\\
\hspace*{0.1in}\\
\underline{Third step: uniqueness of the solution to (\ref{f}) and stability results.}
\hspace*{0.1in}\\
\hspace*{0.1in}\\
The previous line of arguments can be followed to obtain that the solution is unique. Namely, assuming the existence of two possibly local solutions $f_1$ and $f_2$ to (\ref{f}) with the same initial datum and bounded energy, Lemma \ref{initial-layer} holds for both solutions. The difference $g= f_1-f_2$ satisfies
\[ \begin{aligned}
&\partial _tg+\bar{v}\cdot \nabla _xg\\
&= \int B(f_1^\prime f_{1*}^\prime -f_2^\prime f_{2*}^\prime )F(f_1)F(f_{1*})dv_*dn -\int B(f_1f_{1*} -f_2f_{2*})F(f_1^\prime )F(f_{1*}^\prime )dv_*dn\\
&+\int Bf_2^\prime f_{2*}^\prime \Big( F(f_{1*})\big( F(f_1)-F(f_2)\big) +F(f_2)\big( F(f_{1*})-F(f_{2*})\big) \Big) dv_*dn\\
& -\int Bf_2f_{2*}\Big( F(f_{1*}^\prime )\big( F(f_1^\prime )-F(f_2^\prime )\big) +F(f_2^\prime )\big( F(f_{1*}^\prime )-F(f_{2*}^\prime )\big) \Big) dv_*dn.
\end{aligned}\]
The first line in the r.h.s. of the former equation gives rise to $c\int \lvert g^\sharp (t,x,v)\rvert dxdv$ in the bound from above of $\frac{d}{dt}\lvert g^\sharp (t,x,v)\rvert dxdv$, whereas the two last lines in the r.h.s of the former equation give rise to the bound $c(1+t^{\alpha -1})\int \lvert g^\sharp (t,x,v)\rvert dxdv$. Consequently,
\begin{eqnarray*}
\frac{d}{dt}\int \lvert g^\sharp (t,x,v)\rvert dxdv\leq c(1+t^{\alpha -1})\int \lvert g^\sharp (t,x,v)\rvert dxdv.
\end{eqnarray*}
This implies that $\int \lvert g^\sharp (t,x,v)\rvert dxdv$ is identically zero, since it is zero initially.\\
\hspace*{0.1in}\\
The proof of stability is similar.\\
\hspace*{0.1in}\\
\hspace*{0.1in}\\
%
%
%
%
\section{Conservations of mass, momentum and energy.}\label{conservations}
\setcounter{theorem}{0}
\setcounter{equation}{0}
The conservation of mass and momentum of $f$ follow from the boundedness of the total energy. The energy is non-increasing by the construction of $f$. Energy conservation will follow if the energy is non-decreasing. This requires the preliminary control of the mass density over large velocities, performed in the following lemma.\\
%
%
\begin{lemma}\label{large-velocity-2}
\hspace*{0.02in}\\
 Given $t\in [ 0,T] $, there is a constant $c_t^\prime >0$ such that for every $\lambda >2$ the solution $f$ of (\ref{f})-(\ref{init}) satisfies
\begin{eqnarray*}
\int_{|v|>\lambda} \sup_{(s,x)\in [ 0,t] \times [0,1]^k}f^\sharp (s,x,v)dv\leq\frac{c'_t}{\sqrt{\lambda}}.
\end{eqnarray*}
\end{lemma}
\underline{Proof of Lemma \ref{large-velocity-2}.}\\
Take $\lambda >2$. First consider the case $k= 1$. It follows from (\ref{control-by-gain}) that  
\begin{eqnarray}
 \int _{|v|>\lambda }\sup_{(s,x)\in [ 0,t] \times [ 0,1] }f^\sharp (s,x,v)dv
\leq  \int _{|v|>\lambda }\sup_{x\in [ 0,1] }f_0(x,v)dv+\parallel F_\alpha \parallel _\infty ^2C,
\end{eqnarray}
where
\begin{eqnarray*}
C= \int _{|v|>\lambda }\sup _{x\in [ 0,1] }\int_0^t\int B
 f^{\#}(s,x+s(v_1-v^\prime _1),v^\prime )f^{\#}(s,x+s(v_1-v^\prime _{*1}),v^\prime _*)dvdv_*dnds.
\end{eqnarray*}
For $v',v'_*$ outside of the angular cutoff {(2.2)}, let $n$ be the unit vector in the direction $v-v'$ and $n_\perp $ its orthogonal unit vector in the direction $v-v^\prime _*$. Split $C$ into $C= \sum _{0\leq i\leq 2}C_i$, where
\begin{eqnarray*}
C_0= \int _{|v|>\lambda }\sup _{x\in [ 0,1] }\Big( \int_0^t\int _{|n_1|<\epsilon \text{  or  } |n_{\perp 1}|<\epsilon }B
 f^{\#}(s,x+s(v_1-v^\prime _1),v^\prime )f^{\#}(s,x+s(v_1-v^\prime _{*1}),v^\prime _*)dv_*dnds\Big) dv,
\end{eqnarray*}
and $C_1$ (resp. $C_2$) refers to integration {with respect to $(v_*,n)$ on}
\begin{align*}
&\{ (v_*,n); \quad |n_1|\geq \epsilon, \quad |n_{\perp 1}|\geq \epsilon, \quad   |v^\prime | \geq |v^\prime _*|\} ,\\
\big( \text{resp.    } &\{ (v_*,n); \quad |n_1|\geq \epsilon, \quad |n_{\perp 1}|\geq \epsilon, \quad   |v^\prime | \leq |v^\prime _*|\} \big) .
\end{align*}
By Lemma \ref{df-T-local} and the change of variables $(v,v_*,n)\rightarrow (v_*,v,n_\perp )$,
\begin{equation}\label{C0}
C_0\leq c\epsilon t,
\end{equation}
for some constant $c>0$. Analogously to the control of $A_1$ in the proof of Lemma \ref{df-T-local} and using Lemma \ref{mass-tails}, it holds that
\begin{align*}
C_1&\leq \int _{|v|\geq \lambda }\sup _{x\in [ 0,1] }\int _{|v^\prime |>|v^\prime _*|}B(\int _0^t\sup _{\tau \in [0,t]}f^\sharp (\tau ,x+s(v_1-v_1^\prime ),v^\prime )ds)\\
&\hspace*{1.65in}\sup _{(\tau ,X)\in [0,t] \times [0,1]}f^\sharp (\tau ,X,v^\prime _*)dvdv_*dn\\
&=  \int _{|v|\geq \lambda }\sup _{x\in [ 0,1] }\int _{|v^\prime |>|v^\prime _*|}\frac{B}{|v_1-v^\prime _1|}(\int _{y\in (x,x+t(v_1-v^\prime _1)}\sup _{\tau \in [0,t]}f^\sharp (\tau ,y,v^\prime )dy)\\
&\hspace*{2.9in}\sup _{(\tau ,X)\in [0,t] \times [0,1]}f^\sharp (\tau ,X,v^\prime _*)dvdv_*dn\\
&\leq \int _{|v|\geq \lambda , |v^\prime |>|v^\prime _*|}B\frac{E(t|v_1-v^\prime _1|)+1}{|v_1-v^\prime _1|}(\int _0^1\sup _{\tau \in [0,t]}f^\sharp (\tau ,y,v^\prime )dy)\\
&\hspace*{2.3in}\sup _{(\tau ,X)\in [0,t] \times [0,1]}f^\sharp (\tau ,X,v^\prime _*)dvdv_*dn\\
&\leq c(t+\frac{1}{\epsilon \gamma \gamma ^\prime })\int _{|v|\geq \frac{\lambda }{\sqrt{2}}}\int _0^1\sup _{\tau \in [0,t]}f^\sharp (\tau ,y,v)dydv\\
&\leq \frac{c}{\lambda }(1+\frac{1}{\epsilon }),\quad t\leq \max \{ 1,T\} .
\end{align*}
The term $C_2$ can be controlled similarly to $C_1$ with the change of variables $s\rightarrow y= x+s(v_1-v^\prime _{*1})$. And so,
\begin{align*}
&C\leq c\big( \epsilon +\frac{1}{\lambda }+\frac{1}{\epsilon \lambda }\big) ,\quad t\leq \max \{ 1,T\} .
\end{align*}
Choosing $\epsilon = \frac{1}{\sqrt{\lambda }}$ leads to
\begin{align*}
&C\leq \frac{c}{\sqrt{\lambda }},\quad t\leq \max \{ 1,T\} .
\end{align*}
\hspace*{0.1in}\\
Repeating the previous proof up to time $T$, the lemma follows. \\
In the case of Theorem 2.2 where in particular $k\in \{ 1, 2, 3\} $ and (\ref{Hyp3-f0}) is assumed, analogously to the proof of Lemma \ref{df-T-global} we obtain
\begin{eqnarray*}
\int \sup_{(s,x)\in [0,t] \times [0,1]^k} |v|^2f(s,x,v)dv\leq \tilde{c}_0e^{ct},
\end{eqnarray*}
for some constant $c$. It follows that
\begin{align*}
\int_{|v|>\lambda} \sup_{(s,x)\in [ 0,t] \times [0,1]^k}f^\sharp (s,x,v)dv&\leq\frac{1}{{\lambda}^2}\int|v|^2 \sup_{(s,x)\in [ 0,t] \times [0,1]^k}f^\sharp (s,x,v)dv\\
&\leq\frac{\tilde{c}_0e^{ct}}{\lambda^2}.
\end{align*}
\cqfd
%
%
%
\begin{lemma}\label{conservation-energy}
The solution $f$ to the Cauchy problem (\ref{f})-(\ref{init}) conserves energy.
\end{lemma}
\underline{Proof of Lemma \ref{conservation-energy}.}\\
It remains to prove that the energy is non-decreasing. Taking $\psi_\epsilon=\frac{|v^2|}{1+\epsilon|v|^2}$  as approximation for $|v|^2$, it is enough to bound
\begin{eqnarray*}
\int Q(f)(t,x,v)\psi_\epsilon (v)dxdv = \int B\psi_{\epsilon}\Big( f^\prime f^\prime _{*}F(f)F(f_{*})
- ff_{*}F(f^\prime )F(f^\prime _{*})\Big) dxdvdv_*dn
\end{eqnarray*}
from below by zero in the limit $\epsilon \rightarrow 0$. Similarly to{ \cite{Lu2}},
\[ \begin{aligned}
\int Q(f)\psi_\epsilon dxdv
&=\frac{1}{2}\int B ff_{*}F(f^\prime )F(f^\prime _{*}\Big( \psi_\epsilon(v')+\psi_\epsilon(v'_*)-\psi_\epsilon(v)-\psi_\epsilon(v_*)
\Big)dxdvdv_*dn\\
&\geq -\int Bff_{*}F(f^\prime )F(f^\prime _{*})\frac{\epsilon |v|^2|v_*|^2}{(1+\epsilon|v|^2)(1+\epsilon|v_*|^2)}dxdvdv_*dn.
\end{aligned}\]
The previous line, with the integral taken over a bounded set in $(v,v_*)$, converges to zero when $\epsilon\rightarrow 0$. In  integrating over $|v|^2+|v_*|^2\geq2\lambda^2$ , there is symmetry between the subset of the domain with $|v|^2>\lambda^2$ and the one with $|v_*|^2>\lambda^2$. We discuss the first sub-domain, for which the integral in the last line is bounded from below by
\[ \begin{aligned}
&-c\int |v_*|^2f(t,x,v_*)dxdv_*\int_{|v|\geq \lambda} B \sup_{(s,x)\in [ 0,t] \times [0,1]^k}f^\#(s,x,v)dvdn\\
&\geq -c\int_{|v|\geq \lambda} \sup_{(s,x)\in [ 0,t] \times [0,1]^k}f^\#(s,x,v)dv.
\end{aligned}\]
It follows from Lemma \ref{large-velocity-2} that the right hand side tends to zero when $\lambda \rightarrow \infty$.\\
This implies that the energy is non-decreasing, and bounded from below by its initial value. \\
That completes the proof of the lemma.     \cqfd
\\
$\bold{Acknowledgement.}$ The authors wish to thank the anonymous referees for several  important suggestions to improve the manuscript.\\
\\
\\

\[\]
Leif Arkeryd, arkeryd@chalmers.se\\
Anne Nouri, anne.nouri@univ-amu.fr\\
\newpage
Running head: On a Boltzmann equation for Haldane statistics.\\
\hspace*{1.in}\\
Name and mailing address of the author to whom proofs should be sent.\\
Anne Nouri, \\
anne.nouri@univ-amu.fr
\end{document}